\documentclass[]{revtex4-1}

\pdfoutput=1

\usepackage{amsmath}
\usepackage{amssymb}
\usepackage[english]{babel}
\usepackage{bm}
\usepackage{graphicx}
\usepackage[table]{xcolor}
\begin{document}

\title{Deep Graphs - a general framework to represent and analyze
heterogeneous complex systems across scales}

\author{Dominik Traxl}
\email{dominik.traxl@posteo.org}
\affiliation{Department of Physics, Humboldt Universit\"at zu Berlin, Germany}
\affiliation{Bernstein Center for Computational Neuroscience, Berlin, Germany}
\affiliation{Potsdam Institute for Climate Impact Research, Potsdam, Germany}

\author{Niklas Boers}
\affiliation{Geosciences Department and Laboratoire de
M\'et\'eorologie Dynamique, Ecole Normale Sup\'erieure, Paris, France}
\affiliation{Potsdam Institute for Climate Impact Research, Potsdam, Germany}

\author{J\"urgen Kurths}
\affiliation{Department of Physics, Humboldt Universit\"at zu Berlin, Germany}
\affiliation{Potsdam Institute for Climate Impact Research, Potsdam, Germany}
\affiliation{Department of Control Theory, Nizhny Novgorod State University,
603950 Nizhny Novgorod, Russia}

\date{\today}

\begin{abstract}

Network theory has proven to be a powerful tool in describing and analyzing
systems by modelling the relations between their constituent objects.
Particularly in recent years, great progress has been made by augmenting
`traditional' network theory in order to account for the multiplex nature of
many networks, multiple types of connections between objects, the time-evolution
of networks, networks of networks and other intricacies. However, existing
network representations still lack crucial features in order to serve as a
general data analysis tool. These include, most importantly, an explicit
association of information with possibly heterogeneous types of objects and
relations, and a conclusive representation of the properties of groups of nodes
as well as the interactions between such groups on different scales. In this
paper, we introduce a collection of definitions resulting in a framework that,
on the one hand, entails and unifies existing network representations (e.g.,
network of networks, multilayer networks), and on the other hand, generalizes
and extends them by incorporating the above features. To implement these
features, we first specify the nodes and edges of a finite graph as sets of
properties (which are permitted to be arbitrary mathematical objects). Second,
the mathematical concept of partition lattices is transferred to network theory
in order to demonstrate how partitioning the node and edge set of a graph into
supernodes and superedges allows to aggregate, compute and allocate information
on and between arbitrary groups of nodes. The derived partition lattice of a
graph, which we denote by \textit{deep graph}, constitutes a concise, yet
comprehensive representation that enables the expression and analysis of
heterogeneous properties, relations and interactions on all scales of a complex
system in a self-contained manner. Furthermore, to be able to utilize existing
network-based methods and models, we derive different representations of
multilayer networks from our framework and demonstrate the advantages of our
representation. On the basis of the formal framework described here, we provide
a rich, fully scalable (and self-explanatory) software package that integrates
into the PyData ecosystem and offers interfaces to popular network packages,
making it a powerful, general-purpose data analysis toolkit. We exemplify an
application of deep graphs using a real world dataset, comprising 16 years of
satellite-derived global precipitation measurements. We deduce a deep graph
representation of these measurements in order to track and investigate local
formations of spatio-temporal clusters of extreme precipitation events.

\end{abstract}

% subject areas: (Interdisciplinary Phyiscs, Complex Systems, Statistical
% Physics)


\maketitle

\tableofcontents

\newpage

The main focus of this paper is to provide a formal framework that enables a
mathematically accurate description of any given system in a self-contained
fashion. In addition, the purpose of this framework is to facilitate the
utilization of existing methods and models supporting a practical data analysis.
Network theory serves as the mathematical foundation of our framework. A network
models the elements of a system as nodes, and their relations (or interactions)
as edges. Particularly in the recent past -- certainly also due to the deluge of
available data -- one could notice a large number of publications attempting to
augment `traditional' networks, in order to accommodate the increased
heterogeneity of data, and to assign labels and values to nodes and edges (e.g.
networks of networks, multilayer networks). The framework proposed here entails
and unifies these approaches, but also generalizes them with two main aspects in
mind: 1. Any node and any edge may be assigned possibly \textit{distinct} types
of properties (e.g., a node representing a human being may have `age' as a type
of property whose value is a number, and `blood values' as another type of
property whose value is a table of labels and numbers). 2. Integration of
properties of groups of nodes and their respective interrelations within the
same framework. Together, these objectives make it possible to combine different
datasets (e.g., climatological and socioecological data or
(electro)physiological records of different organs), integrate a priori
knowledge of groups of objects and their relations, and carry out an analysis of
potential relationships of the respective systems within the same network
representation. On the basis of the mathematical work we provide here, existing
network measures can be generalized and new measures developed. Yet, in order to
practically conduct data analysis, we also provide a rich software
implementation of our framework that integrates into the PyData ecosystem (which
is comprised of various libraries for scientific computing), and offers
interfacing methods to popular network packages, making it a considerable
general-purpose data analysis toolkit.

\section{Introduction}

At the present time, we are observing a quantification of our world at an
unprecedented rate~\cite{90perc}. On the one hand -- due to the rapid
technological progress -- we are extracting an ever increasing amount of
information from nature, ranging from subatomic to astronomical scales. On the
other hand, we are producing a vast amount of information in our daily lives
interacting with electronic devices, thereby generating traceable information,
tracked and stored by us personally, but also by organizations, companies and
governments.

From a scientific point of view, this rapid increase in the amount and
heterogeneity of available data poses both a great opportunity, but also
methodological challenges: how can we describe and represent complex systems,
made of multifarious subsystems interacting intricately on various scales; and
once we have a suitable representation, how do we detect patterns and
correlations, develop and test hypotheses and eventually come up with models and
working theories of underlying mechanisms?

Thankfully, we can look back on centuries of scientific progress, tackling these
questions. Rich tool sets to represent, analyze and model systems have been
developed in various fields, such as: probability theory~\cite{Jaynes2005};
multivariate statistics~\cite{anderson2003introduction}; non-linear
dynamics~\cite{Strogatz1994, thiel2010nonlinear}; game
theory~\cite{Osborne1994}; graph theory~\cite{Newman2010}; or machine
learning~\cite{hastie2009elements, bishop2006pattern, haykin2009neural,
vapnik1998statistical, deeplearning}.

In this paper, we propose a framework that is capable of representing
arbitrarily complex systems in a self-contained manner, and establishes an
interface for the tools and methods developed in research disciplines such as
those mentioned above. The framework is based on the ontological assumption that
every system can be described in terms of its constituent objects (anything
conceivable, i.e., ``beings'', ``things'', ``entities'', ``events'', ``agents'',
``concepts'' or ``ideas'') and their relations. With this assumption in mind, we
build this framework based on graph or network theory. A graph, in its simplest
form, is a collection of nodes (representing objects) where some pairs of nodes
are connected by edges (representing the existence of a
relation)~\cite{Bollobas1998}. On top of that, we define an additional structure
in order to meet the following objectives:
\begin{enumerate}
  \item  any node of the network may explicitly incorporate properties of the
object(s) it represents. We refer to these properties as the \textit{features}
of a node, which themselves are mathematical objects.

  \item any edge of the network may explicitly incorporate properties of the
relation(s) it represents. We refer to these properties as the
\textit{relations} of an edge, which themselves are mathematical objects.

  \item any subset of the set of all nodes of the network may be grouped into a
\textit{supernode}. Thereby, we may aggregate the features of the supernodes'
constituent nodes. Furthermore, we may allocate features particular to that
supernode (``emergent'' properties of the compound supernode), based on either
the aggregated features, a priori knowledge, or both.

  \item any subset of edges of the set of all edges of the network may be
grouped into a \textit{superedge}. Thereby, we may aggregate the relations of
the superedges' constituent edges. Furthermore, we may allocate relations
particular to that superedge (``emergent'' properties of the compound
superedge), based on either the aggregated relations, a priori knowledge, or
both.

  \item we may place edges between any pair of supernodes, as well as between
supernodes and nodes.
\end{enumerate}
We believe that a comprehensive treatment of groups of objects, as well as their
relations, is just as indispensable as an explicit incorporation of data, not
only in the representation of complex systems, but also in their analysis.
First, because it facilitates the means to represent features, relations and
interactions on different scales, and second, because it allows us to
coarse-grain, simplify and highlight important large-scale structures in a
data-driven analysis.

Needless to say, this is not the first attempt to augment simple graphs in order
to satisfy at least some of the above objectives. In weighted graphs, for
instance, one can assign a number to each edge (i.e., the weight, strength, or
distance of an edge)~\cite{horvath2014weighted}. In node-weighted networks, it
is possible to assign numbers to the nodes of the network~\cite{Wiedermann2013}.
In hypergraphs, one can define edges joining more than two vertices at a time
(called hyperedges), essentially allowing for the assignment of groups in a
network~\cite{berge1976graphs}. Such a membership of nodes in groups can also be
represented by bipartite networks, where one of two kinds of nodes represents
the original objects, and the other kind represents the groups to which the
objects belong~\cite{asratian1998bipartite}. Particularly in recent years -- due
to the deluge of available data -- a multitude of frameworks has been proposed,
with the aim of pluralizing the number of labels and values that may be assigned
to a node, and allowing for different categories of connections between pairs of
nodes, such as, e.g.: multivariate networks; multidimensional networks;
interacting networks; interdependent networks; networks of networks;
heterogeneous information networks; and multilayer networks (see
\cite{boccaletti2014structure, kivela2014multilayer, Berlingerio2011, Han2012,
DeDomenico2013,Gao2011a,Gao2011b,santiago2008extended} and references therein).

However, none of these frameworks satisfies all of the above objectives at the
same time. In contrast, the framework proposed in this paper meets all these
objectives. This allows us, on the one hand, to derive all of the above network
representations as special cases by imposing certain constraints on our
framework, which enables the utilization of the network-based methods, models
and measures developed for them. On the other hand, we will demonstrate how the
implementation of these objectives into our framework generalizes existing
network representations, making it possible to combine heterogeneous datasets,
integrate a priori knowledge of groups of objects and their relations, and
conduct an analysis of potential interrelations of the respective systems within
the same network representation. Considering the theoretical work provided here,
existing network-measures may be generalized and new measures developed,
particularly in respect of the heterogeneity of a system's components and their
interactions on different scales. Based on the introduced framework, we also
provide a general-purpose data analysis software package~\cite{deepgraph} that
is fully scalable and integrates into the PyData ecosystem comprised of various
libraries for scientific computing~\cite{pydata}. Apart from providing its own
graph-theoretic data structure to accommodate the above objectives, our software
package also provides interfaces for known data structures such as adjacency
lists, adjacency matrices, incidence matrices and tensors, which have recently
been introduced to represent multilayer networks~\cite{DeDomenico2013}.

The paper is structured as follows: the theoretical part of our framework is
described in Sec.~\ref{sec:Graph Representation}, where we introduce our
representation of a graph, and Sec.~\ref{sec:Graph Partitioning}, where we
demonstrate a comprehensive manner of graph partitioning. Thereafter, we outline
the general procedure of constructing a deep graph and demonstrate how our
framework integrates with existing data analysis tools in Sec.~\ref{sec:Deep
Graph Construction}. We then demonstrate a real world application of our
framework on a global precipitation dataset in Sec.~\ref{sec:Application to
Global Precipitation Data}, before we draw our conclusions in
Sec.~\ref{sec:Conclusion}.

\section{Graph Representation}\label{sec:Graph Representation}

Throughout this paper, we assume (w.l.o.g.) that (super)nodes, (super)edges,
types of features and types of relations are represented by consecutive integers
starting from 1. Also, there is a glossary in Tab.~\ref{tab:glossary}, which
summarizes all the important quantities of a deep graph.

The basis of our representation is a finite, directed graph (possibly with self
loops), given by a pair
\begin{equation}
G = (V, E),
\end{equation}
where $V$ is a set of $n := |V|$ nodes,
\begin{equation}
V = \lbrace V_i \;\vline\; i \in \lbrace 1, 2, ..., n \rbrace \rbrace,
\end{equation}
and $E$ is a set of $m := |E|$ directed edges, given by
\begin{equation}
E \subseteq \lbrace E_{ij} \;\vline\; i,j \in \lbrace 1, 2, ..., n \rbrace
\rbrace =: E'.
\end{equation}
Every node $V_i \in V$ of this graph represents some object(s), and every edge
$E_{ij} \in E$ represents the existence of some relation(s) from node $V_i$ to
node $V_j$. We say that an edge $E_{ij}$ is \textit{incident} to both nodes
$V_i$ and $V_j$. In order to explicitly incorporate information or data of the
objects and their pairwise relations, we specify every node $V_i$ and every edge
$E_{ij}$ of $G$ as a set of its respective properties. We refer to the
properties of a node as its \textit{features}, and to the properties of an edge
as its \textit{relations}.

\begin{figure*}
\includegraphics[width=1\linewidth]{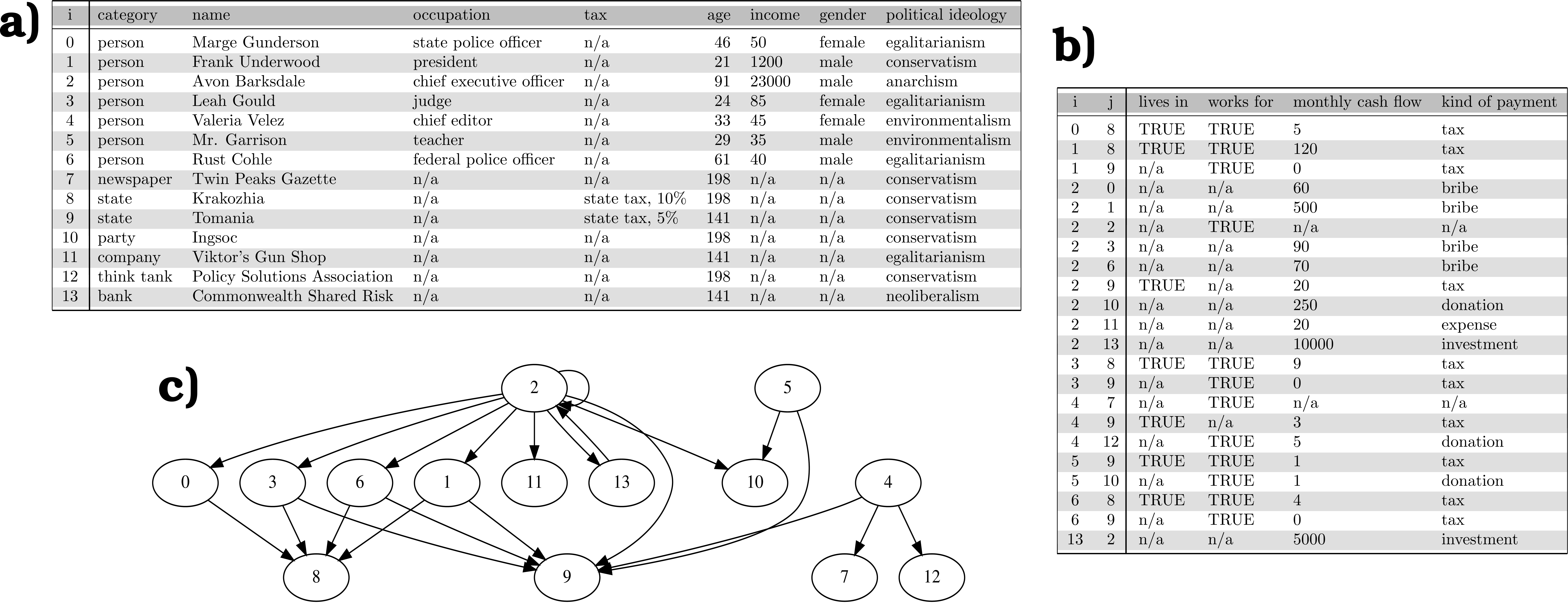}
\caption{\label{fig:example_graph} \textsf{Graph Representation. Illustration
of a fictional graph $G = (V, E)$ consisting of $n=14$ nodes and $m=22$
directed edges.} \textbf{(a)} Representation of the nodes $V_i \in V$. The left
column indicates the nodes' indices, the top row indicates the \textit{types}
of features. A feature denoted ``n/a'' means that the corresponding node does
not have a feature of the corresponding type. \textbf{(b)} Representation of
the edges $E_{ij} \in E$. The first two columns from the left indicate the
indices of edges $E_{ij}$ from node $V_i$ to node $V_j$ and the top row
indicates the \textit{types} of relations. A relation denoted ``n/a'' means
that the corresponding edge does not have a relation of the corresponding type.
\textbf{(c)} Depiction of the graph's topology, where nodes are represented by
indexed circles and edges are represented by arrows.}
\end{figure*}

Hence, we define every node $V_i$ as a set of $f_i$ features (and its index, to
guarantee uniqueness of the nodes), given by
\begin{equation}
V_i = \lbrace i, F_i^1, F_i^2, ..., F_i^{f_i} \rbrace.
\end{equation}
As opposed to the `weight' of a node in node-weighted
networks~\cite{Wiedermann2013} -- which is usually a real number -- a feature
$F_i^j$ can be any mathematical object (e.g. numbers; quantitative or
categorical variables; sets; matrices; tensors; functions; nodes; edges; graphs;
but also strings to represent abstract objects, such as concepts or ideas).
Furthermore, we associate every feature with a \textit{type}, in order to
express the kind of property a feature is related to and to establish a
comparability between the features of different nodes. For example, for a node
representing a city, some types of features might be `location', `age', `number
of inhabitants', `unemployment rate' and `voting patterns'. For a node
representing a neuron, some types of features could be `time series of the
membrane potential', `measuring device' and `distribution of ion channel types'.
On that account, we denote with $F = \lbrace F^j_i \;\vline\; i \in \lbrace 1,
2, ..., n \rbrace \land j \in \lbrace 1, 2, ..., f_i \rbrace \rbrace$ the set of
all features, and with $T_v = \lbrace 1, 2, ..., n_{\text{types}} \rbrace$ the
set of all distinct types of features contained in the graph $G$. We then define
a surjective function mapping every feature to its corresponding type,
\begin{equation}
t_v: F \rightarrow T_v,  F^j_i \mapsto t_v(F^j_i) := T^j_i \in T_v,
\end{equation}
such that $t_v(F^j_i) = t_v(F^l_k)$ for all pairs of features that share the
same type.
However, we do not allow a node $V_i$ to have multiple features of the same
type, $t_v(F^j_i) \neq t_v(F^k_i)$ for all $j \neq k \in \lbrace 1, 2, ..., f_i
\rbrace$. In other words, every node $V_i$ has exactly $f_i$ distinct types of
features. Figure~\ref{fig:example_graph}(a) depicts different nodes along with
their features and the feature's types.

Analogously, we define every edge $E_{ij} \in E$ as a set of $r_{ij}$ relations
(and its index pair, to guarantee uniqueness of the edges), given by
\begin{equation}
E_{ij} = \lbrace (i, j), R_{ij}^1, R_{ij}^2, ..., R_{ij}^{r_{ij}} \rbrace.
\end{equation}
Again, as opposed to the commonly real-valued `weight' of an edge in
edge-weighted networks~\cite{horvath2014weighted}, a relation $R_{ij}^k$ can be
any mathematical object. Just like features, we map every relation to its
corresponding type, indicating the kind of property of a relation (e.g.
`distance between', `correlation between', `similarity between', `works for',
`is part of'). We denote with $R = \lbrace R^k_{ij} \;\vline\; i,j \in \lbrace
1, 2, ..., n \rbrace \land k \in \lbrace 1, 2, ..., r_{ij} \rbrace \rbrace$ the
set of all relations, and with $T_e = \lbrace 1, 2, ..., m_{\text{types}}
\rbrace$ the set of all distinct types of relations contained in the graph $G$.
We then map every relation onto a type,
\begin{equation}
t_e: R \rightarrow T_e, R^k_{ij} \mapsto t_e(R^k_{ij}) := T^k_{ij} \in T_e,
\end{equation}
such that $t_e(R^k_{ij}) = t_e(R^l_{mn})$ for all pairs of relations that share
the same type, and $t_e(R^k_{ij}) \neq t_e(R^l_{ij})$ for all $k \neq l \in
\lbrace 1, 2, ..., r_{ij} \rbrace$. Therefore, every edge $E_{ij}$ has exactly
$r_{ij}$ distinct types of relations. Figure~\ref{fig:example_graph}(b)
illustrates several edges with different types of relations.

For notational convenience later on, we define all elements $E_{ij} \in E'$ that
are not in $E$ as empty sets,
\begin{equation}
E_{ij} := \varnothing \text{ for all } i, j \in
\lbrace 1, 2, ..., n \rbrace: E_{ij} \in E' \setminus E.
\end{equation}
In other words, we say an edge from node $V_i$ to node $V_j$ exists if $E_{ij}
\neq \varnothing$, and in this context, we term $V_i$ the \textit{source} node
and $V_j$ the \textit{target} node. Therefore, we can rewrite the set of edges
of $G = (V, E)$ as
\begin{equation}
E = \lbrace E_{ij} \;\vline\; i,j \in \lbrace 1, 2, ..., n \rbrace \land
E_{ij} \neq \varnothing \rbrace.
\end{equation}

\section{Graph Partitioning} \label{sec:Graph Partitioning}

In this section, we introduce a comprehensive concept of graph
partitioning. To avoid confusion: we do not refer to graph partitioning in the
sense of finding ``good'' partitions (i.e. communities) based on some cost
function or statistical measures~\cite{Buluc2013} such as, e.g., Newman's
modularity measure~\cite{Newman2004}. Instead, we refer to graph partitioning in
the more general sense of partitions of sets~\cite{Lucas1990}.

First, we demonstrate how partitioning the node set $V$ of a graph $G = (V, E)$
enables us to group arbitrary nodes into \textit{supernodes}, and equivalently,
how partitioning the edge set $E$ allows us to group arbitrary edges into
\textit{superedges}. Then, we introduce a coherent manner of partitioning a
graph $G = (V, E)$ into a \textit{supergraph}, where the edge set $E$ is
partitioned in accordance with a given partition of the node set $V$, based on
the edges' incidences to the nodes.

Partitioning nodes, edges or graphs -- as we will show -- not only conserves the
information contained in the graph $G = (V, E)$, but allows us to redistribute
it. This enables us to aggregate the features and relations of any desirable
group of nodes and edges, and to allocate information particular to them.
Furthermore, it facilitates the means to place edges between any supernodes or
between supernodes and nodes, allowing us to represent interactions or relations
on any scale of a complex system.

\subsection{Partitioning Nodes} \label{sec:Partitioning Nodes}

Given a graph $G = (V,E)$ with $n = |V|$ nodes, we define a surjective
function mapping every node $V_i \in V$ to a supernode label (i.e. feature)
${}^vS_i$,
\begin{equation}
{}^vp: V \rightarrow {}^vS = \lbrace  1, 2, ..., n^p \rbrace, V_i \mapsto
{}^vp(V_i) := {}^vS_i \in {}^vS.
\label{eq:p_nodes}
\end{equation}
This function induces a partition $V^p$ of $V$ into $n^p = |V^p|$ supernodes
$V^p_i$, given by
\begin{align}
&V^p_i = \lbrace V_j \;\vline\; j \in \lbrace 1, 2, ..., n \rbrace \wedge
{}^vp(V_j) = {}^vS_i \rbrace, \text{ and} \label{eq:np1} \\
&V^p = \lbrace V^p_i \;\vline\; i \in \lbrace 1, 2, ..., n^p \rbrace \rbrace.
\label{eq:np2}
\end{align}
The number of nodes a supernode $V^p_i \in V^p$ contains is denoted by $n^{p,i}
:= |V^p_i|  \geq 1$.

The supernode labels given by the function ${}^vp(V_i) = {}^vS_i$ can be
transferred as features to the nodes of $G$,
\begin{equation}
V_i = \lbrace i, F^1_i, F^2_i, ..., F^{f_i}_i, {}^vS_i \rbrace,
\end{equation}
where the type of feature of ${}^vS_i$ is
the same for all nodes, $t_v({}^vS_i) = t_v({}^vS_j)$ for all $i, j \in \lbrace
1, 2, ..., n \rbrace$. In turn, every feature itself can be interpreted as a
supernode label, and we can say that its corresponding type induces a partition
of the node set. For instance, looking at Fig.~\ref{fig:example_graph}(a), we
see that the type of feature `political ideology' induces a partition of $V$
into $n^p = 5$ supernodes: `egalitarianism' (consisting of $n^{p,1} = 4$
nodes), `conservatism' ($n^{p,2} = 6$ nodes), `anarchism' ($n^{p,3} = 1$
node), `environmentalism' ($n^{p,4} = 2$ nodes) and `neoliberalism'
($n^{p,5} = 1$ node). Since some nodes might not have a feature of a certain
type [see for instance the type `gender' in Fig.~\ref{fig:example_graph}(a)],
there is a degree of freedom when partitioning by that type: we can create one
supernode comprising all nodes without the feature; create a separate supernode
for every node without the feature; or create no supernode at all for these
nodes. This choice is of course dependent on the analysis.

\subsection{Partitioning Edges} \label{sec:Partitioning Edges}

Partitioning the edge set $E$ of a given graph $G = (V, E)$ with $n = |V|$ nodes
and $m = |E|$ edges can be realized just like partitioning the node set.
However, since edges $E_{ij}$ are incident to pairs of nodes $(V_i, V_j)$, we
later demonstrate how to exploit this association in order to partition edges
based on properties of the nodes. Here, we demonstrate the procedure analogous
to that of partitioning nodes. Hence, we define a surjective function mapping
every edge $E_{ij} \in E$ to a superedge label (i.e. relation) ${}^eS_r$, given
by
\begin{equation}
{}^ep: E \rightarrow {}^eS = \lbrace  1, 2, ..., m^p \rbrace, E_{ij} \mapsto
{}^ep(E_{ij}) := {}^eS_r \in {}^eS.
\label{eq:p_edges}
\end{equation}
This function induces a partition $E^p$ of $E$ into $m^p = |E^p|$ superedges
$E^p_r$, where
\begin{align}
&E^p_r = \lbrace E_{uv} \;\vline\; \Phi^e(u, v) \land {}^ep(E_{uv}) = {}^eS_r
\rbrace, \label{eq:ep1} \\
&\Phi^e(u, v) : (u, v \in \lbrace 1, 2, ..., n \rbrace \land E_{uv}
\neq \varnothing), \text{ and} \label{eq:ep2} \\
&E^p = \lbrace E^p_r \;\vline\; r \in \lbrace 1, 2, ..., m^p \rbrace \rbrace.
\label{eq:ep3}
\end{align}
The number of edges a superedge $E^p_r \in E^p$ contains is denoted by  $m^{p,r}
:= |E^p_r|  \geq 1$.

Equivalently to supernode labels, we can transfer the superedge labels given by
the function ${}^ep(E_{ij}) = {}^eS_r$ as relations to the edges of $G$,
\begin{equation}
E_{ij} = \lbrace (i, j), R^1_{ij}, R^2_{ij}, ..., R^{r_{ij}}_{ij}, {}^eS_r
\rbrace,
\end{equation}
where the type of relation of ${}^eS_r$ is the same for all edges, $t_e({}^eS_i)
= t_e({}^eS_j)$ for all $i, j \in \lbrace 1, 2, ..., m \rbrace$.
Again, every relation itself can be interpreted as a superedge label, and we say
that its corresponding type induces a partition of the edge set. Looking at
Fig.~\ref{fig:example_graph}(b), we see that the type of relation `kind of
payment' induces a partition of $E$ into $m^p = 6$ superedges: `bribe'
(consisting of $m^{p,1} = 4$ edges), `donation' ($m^{p,2} = 3$ edges), `expense'
($m^{p,3} = 1$ edge), `investment' ($m^{p,4} = 2$ edges), `tax' ($m^{p,5} = 10$
edges), and `n/a' ($m^{p,6} = 2$ edges). Since the last two edges do not have a
relation of the type `kind of payment', we could have also partitioned the edges
into $m^p = 5$ superedges (leaving the two edges out), or into $m^p = 7$
superedges (the two edges are put into separate superedges).

\subsection{Partitioning a Graph} \label{sec:Partitioning a Graph}

\begin{figure}
\includegraphics[width=.5\linewidth]{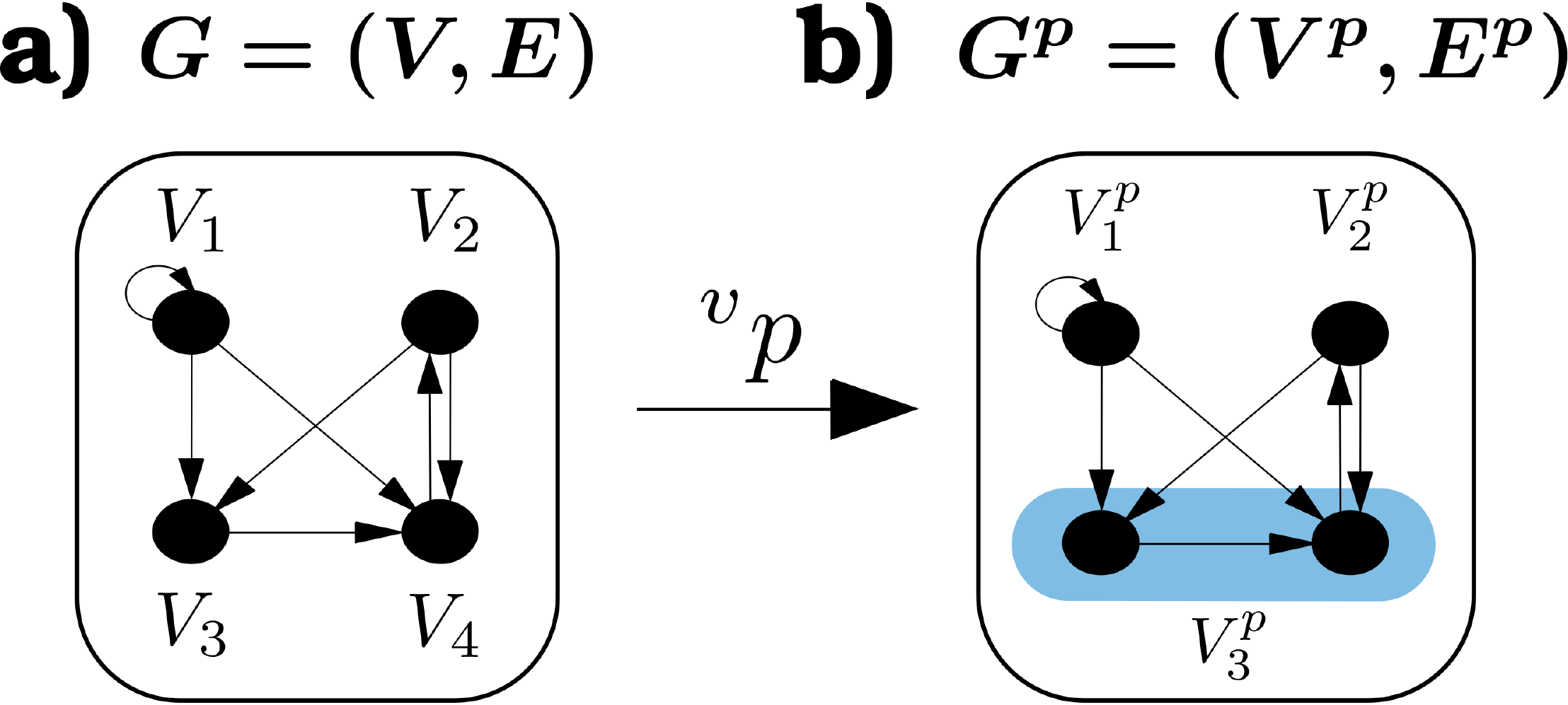}
\caption{\label{fig:natural_supergraph}\textsf{Graph Partitioning. Illustration
of a supergraph, `naturally' induced by a partition of the node set.}
\textbf{(a)} The graph $G = (V, E)$, comprised of $n = 4$ nodes $V = \lbrace
V_1, V_2, V_3, V_4 \rbrace$ and $m = 7$ edges $E = \lbrace E_{11}, E_{13},
E_{14}, E_{23}, E_{24}, E_{34}, E_{42} \rbrace$. \textbf{(b)} The supergraph
$G^p = (V^p, E^p)$, obtained by grouping the nodes $V_3$ and $V_4$ into the
supernode $V^p_3 = \lbrace V_3, V_4 \rbrace$. It is comprised of $n^p = 3$
nodes $V^p = \lbrace V^p_1, V^p_2, V^p_3\rbrace$, and $m^p = 5$ edges $E^p$,
given by: $E^p_{11} = \lbrace E_{11} \rbrace, E^p_{13} = \lbrace E_{13}, E_{14}
\rbrace, E^p_{23} = \lbrace E_{23}, E_{24} \rbrace, E^p_{32} = \lbrace E_{42}
\rbrace, \text{ and } E^p_{33} = \lbrace E_{34} \rbrace$.}
\end{figure}

Here, we introduce a coherent manner of partitioning a graph $G = (V, E)$ with
$n = |V|$ nodes and $m = |E|$ edges, based on the edges' incidences to the
nodes. Given a partition $V^p$ of $V$ induced by a function ${}^vp(V_i) =
{}^vS_i$ [see Eqs.~(\ref{eq:p_nodes})-(\ref{eq:np2})], we define the
\textit{corresponding} partition $E^p$ of $E$ into $m^p = |E^p|$ superedges
$E^p_{ij}$ by the following equations:
\begin{equation}
E^p_{ij} := \lbrace E_{uv} \;\vline\; \Phi^e(u, v) \land {}^vp(V_u) = {}^vS_i
\land {}^vp(V_v) = {}^vS_j \rbrace, \label{eq:gp1}
\end{equation}
where
\begin{align}
&\Phi^e(u,v): \left( u,v \in \lbrace 1, 2, ..., n \rbrace \land E_{uv} \neq
\varnothing  \right), \text{ and} \label{eq:gp2} \\
&E^p := \lbrace E^p_{ij} \;\vline\; i,j \in \lbrace 1, 2, ..., n^p \rbrace
\land E^p_{ij} \neq \varnothing \rbrace. \label{eq:gp3}
\end{align}
By this definition, we group all edges $E_{ij}$ originating from nodes in
supernode $V^p_i$ and targeting nodes in supernode $V^p_j$ into a superedge
$E^p_{ij}$, consisting of $m^{p,ij} := |E^p_{ij}| \geq 0$ edges. It is
straightforward to show that this corresponding partition is indeed a partition
of $E$, and therefore we can say that partitioning the node set $V$ by ${}^vp$
induces a \textit{supergraph} $G^p = (V^p, E^p)$. In reference to the graph in
Fig.~\ref{fig:example_graph}, a partition of the nodes by the type of feature
`category', for instance, would yield a supergraph consisting of $n^p = 7$
supernodes: `bank' (consisting of $n^{p,1} = 1$ node), `company' ($n^{p,2} =
1$ node), `newspaper' ($n^{p,3} = 1$ node), `party' ($n^{p,4} = 1$ node),
`person' ($n^{p,5} = 7$ nodes), `state' ($n^{p,6} = 2$ nodes) and `think
tank' ($n^{p,7} = 1$ node); and $m^p = 8$ corresponding superedges: from
`bank' to `person' (consisting of $m^{p,15} = 1$ edge); and from `person'
to: `bank' ($m^{p,51}=1$ edge), `company' ($m^{p,52}=1$ edge), `newspaper'
($m^{p,53}=1$ edge), `party' ($m^{p,54}=2$ edges), `person' ($m^{p,55}=5$
edges), `state' ($m^{p,56}=10$ edges) and `think tank' ($m^{p,57}=1$ edge).
See also Fig.~\ref{fig:natural_supergraph} for an illustration of grouping a
graph's nodes and edges into a supergraph.

\subsection{The Partition Lattices of a Graph}

In this section, we explain some general mathematical properties that arise when
partitioning a graph. This provides for a deeper understanding of this
framework, and sets the stage for the next sections.

Before we go into details of graph-specific partitioning, we point out some
relevant properties of what in mathematics is known as \textit{partition
lattices}~\cite{Birkhoff1940}. Assume we are given a finite, non-empty
$n$-element set $X$. The total number of distinct partitions we can create of it
is given by the Bell number $B(n)$~\cite{Bell1938,Becker1948}. The set of all
possible partitions, which we denote by $P = \lbrace P_i \;\vline\; i \in
\lbrace 1, 2, ..., B(n) \rbrace \rbrace$, is a \textit{partially ordered} set,
since some of the elements of $P$ have a pair-wise relation, which is called the
\textit{finer-than} relation. A partition $P_i$ is said to be a
\textit{refinement} of a partition $P_j$, if every element of $P_i$ is a subset
of some element of $P_j$. If this condition is fulfilled, one says that $P_i$ is
\textit{finer} than $P_j$, $P_i \leq P_j$, and vice versa, $P_j$ is
\textit{coarser} than $P_i$, $P_j \geq P_i$. Since $X$ is finite, every
partition $P_i$ is bounded from below and from above with respect to this
finer-than relation,
\begin{equation}
P_f \leq P_i \leq P_c, \text{ for all } i \in \lbrace 1,
2, ..., B(n) \rbrace,
\end{equation}
where $P_f$ is called the \textit{finest} element of P,
given by $P_f = \lbrace \lbrace X_1 \rbrace, \lbrace X_2 \rbrace, ..., \lbrace
X_n \rbrace \rbrace$, and $P_c$ is the \textit{coarsest} element, given by the
trivial partition $P_c = \lbrace X \rbrace$. This implies that each set of
elements of $P$ has a finest upper bound and a coarsest lower bound. Therefore,
the set of all possible partitions $P$ is called a partition lattice (or more
precisely, a \textit{geometric} lattice, since $X$ is
finite~\cite{welsh2010matroid}). Any \textit{totally} ordered subset of $P$ is
called a \textit{chain}, and any subset of $P$ for which there exists no
relation between any two different elements of that subset is called an
\textit{antichain}.

Since in this paper we are dealing with finite graphs exclusively, we can
directly build the lattices of the node set $V$ and the edge set $E$, and
translate the above properties of lattices into the context of graphs. However,
we will also make use of the natural way of partitioning a graph as demonstrated
in Sec.~\ref{sec:Partitioning a Graph}, in order to create the geometric lattice
of a graph $G = (V, E)$. This lattice, by construction, entails the lattice of
the node set $V$, and a specific subset of the lattice of the edge set $E$, and
there are therefore two lattices of interest: the lattice of a graph $G$, and
the lattice of its edges $E$.

\begin{figure}
\includegraphics[width=.5\linewidth]{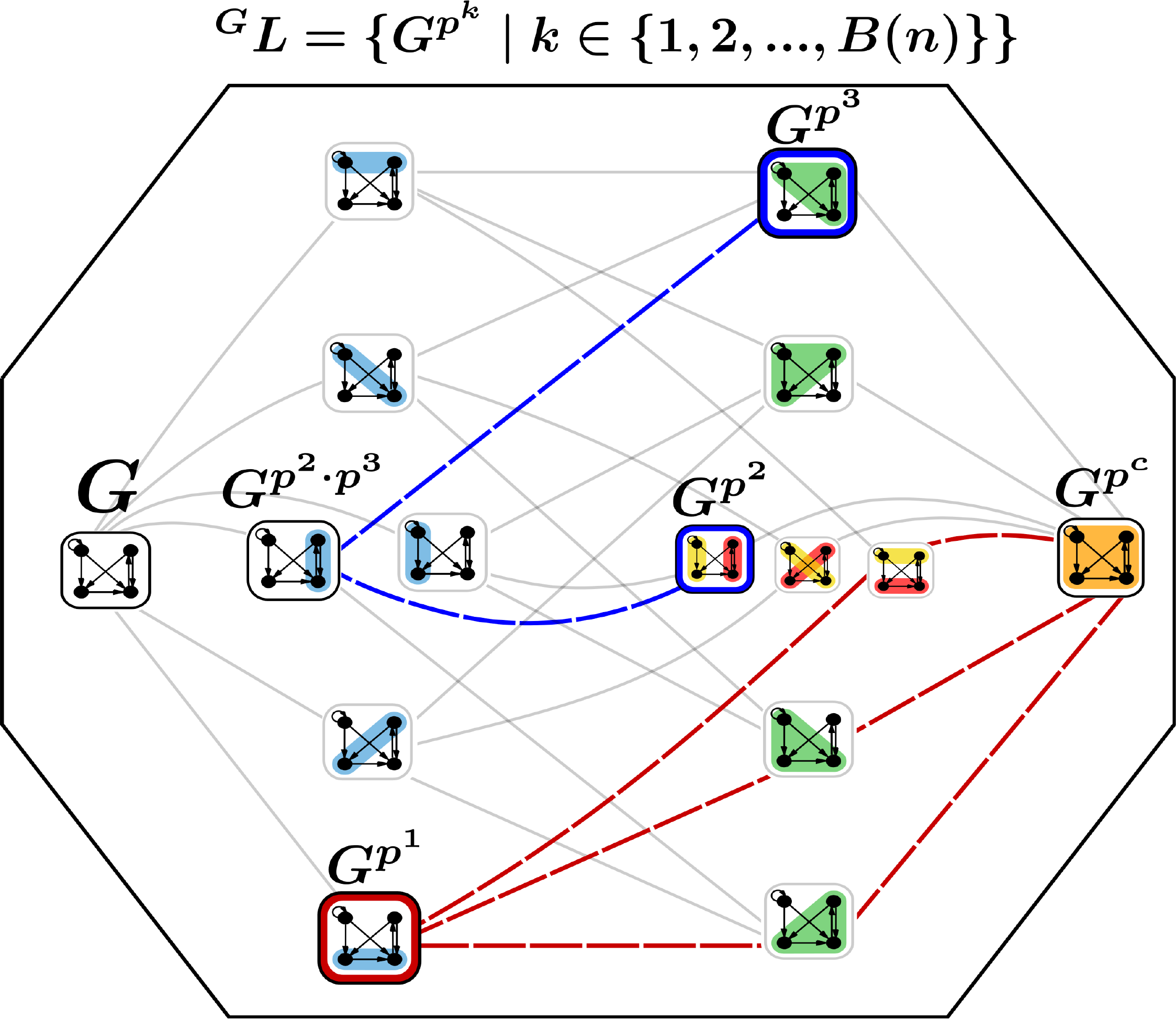} \caption{\label{fig:supergraph_lattice}\textsf{A deep graph, i.e. the
geometric partition lattice of a graph.} \textbf{(a)} Illustration of the graph
$G = (V, E)$ as described in Fig.~\ref{fig:natural_supergraph}(a) (on the very
left of the plot), and its $B(n) = 15$ corresponding supergraphs, ordered by
refinement from the right to the left. The supergraph $G^{p^1}$ is illustrated
in detail in Fig.~\ref{fig:natural_supergraph}(b). Each link in this Hasse
diagram corresponds to the finer-than relation between a pair of supergraphs.
The dashed lines colored in red correspond to chains in the lattice.
\textbf{(b)} An intersection partition is illustrated by $G^{p^2 \cdot p^3}$,
which results from intersecting $G^{p^2}$ and $G^{p^3}$. It constitutes a
refinement of both $G^{p^2}$ and $G^{p^3}$ (blue dashed lines). The figure is a
modification of~\cite{lattice_pic}.}
\end{figure}

Let us note down the lattice of a graph $G$ with $n$ nodes and $m$ edges, for
which there is a total of $B(n)$ different supergraphs. We create the set of all
distinct partitions of $V$ by prescribing a set of functions ${}^vp = \lbrace
{}^vp^k \;\vline\; k \in \lbrace 1, 2, ..., B(n) \rbrace \rbrace$, such that
each function
\begin{align}
&{}^vp^k: V \rightarrow {}^vS^k = \lbrace  1, 2, ..., n^{{}p^k} \rbrace, \\
&V_i \mapsto {}^vp^k(V_i) := {}^vS^k_i \in {}^vS^k,  k \in \lbrace 1, 2, ...,
B(n) \rbrace,
\end{align}
induces a supergraph $G^{p^k} = (V^{p^k}, E^{p^k})$ as demonstrated in
Sec.~\ref{sec:Partitioning a Graph} and illustrated in
Fig.~\ref{fig:natural_supergraph}. The partition lattice of $V$, induced by the
set of functions ${}^vp$, is therefore given by ${}^VL = \lbrace V^{p^k}
\;\vline\; k \in \lbrace 1, 2, ..., B(n) \rbrace \rbrace$. The finer-than
relation between partitions translated to the lattice of $V$ means that if
$V^{p^k} \leq V^{p^l}$, then every supernode $V^{p^l}_i$ of $V^{p^l}$ is the
union of supernodes $V^{p^k}_j \in V^{p^k}$. We transfer the finer-than relation
to graphs, by saying that $G^{p^k} \leq G^{p^l}$, if both $V^{p^k} \leq V^{p^l}$
and $E^{p^k} \leq E^{p^l}$. With reference to
Eqs.~(\ref{eq:gp1})-(\ref{eq:gp3}), we see that for all partitions $V^{p^k} \leq
V^{p^l}$, it follows that $E^{p^k} \leq E^{p^l}$ by construction, and
consequently, we denote with ${}^GL = \lbrace G^{p^k} \;\vline\; k \in \lbrace
1, 2, ..., B(n) \rbrace$ the partition lattice of $G$, henceforth referred to as
the \textit{deep graph} of $G$. The lattice of the graph depicted in
Fig.~\ref{fig:natural_supergraph}(a) is illustrated in
Fig.~\ref{fig:supergraph_lattice}(a). Some of its properties are: the finest
element of ${}^GL$ is the graph $G = (V, E)$ itself; the coarsest element, which
we denote by $G^{p^c} = (V^{p^c}, E^{p^c})$, consists of one supernode connected
to itself by a single superedge; and every chain in ${}^GL$, illustrated by the
red, dashed lines in Fig.~\ref{fig:supergraph_lattice}(a), corresponds to some
agglomerative, hierarchical clustering of the nodes of $G$.
The dashed blue lines in Fig.~\ref{fig:supergraph_lattice}(b) will be explained
in the next section.

However, the lattice of $E$ is generally not covered entirely by the lattice of
$G$. In fact, maximally $B(n)$ of $B(m)$ possible partitions of $E$ are
contained in ${}^GL$, due to the partitioning of $E$ by correspondence [see
Eqs.~(\ref{eq:gp1})-(\ref{eq:gp3})]. The full lattice of $E$ can be created
analogously to that of $V$, by prescribing a set of functions ${}^ep = \lbrace
{}^ep^k \;\vline\; k \in \lbrace 1, 2, ..., B(m) \rbrace \rbrace$, such that
each function
\begin{align}
&{}^ep^k: E \rightarrow {}^eS^k = \lbrace  1, 2, ..., m^{p^k} \rbrace, \\
&E_{ij} \mapsto {}^ep^k(E_{ij}) := {}^eS^k_r \in {}^eS^k, k \in \lbrace 1, 2,
..., B(m) \rbrace,
\end{align}
induces a partition $E^{p^k}$ of $E$ as demonstrated in
Eqs.~(\ref{eq:p_edges})-(\ref{eq:ep3}). The lattice of $E$ is then given by
${}^EL = \lbrace E^{p^k} \;\vline\; k \in \lbrace 1, 2, ..., B(m) \rbrace
\rbrace$.

In the next section, we introduce a useful tool that can be used to navigate the
lattices ${}^GL$ and ${}^EL$ for the sake of creating meaningful partitions,
based on the features and relations of a given graph.

\subsection{Intersection Partitions} \label{sec:Intersection Partitions}

Due to the rapid increase of possible partitions with growing numbers of nodes
and edges, it is only possible to actually compute the full lattices of $G$ and
$E$ for very small graphs. However, we are generally not interested in every
single partition, but rather a meaningful subset of them. Here, we demonstrate
how to create \textit{intersection partitions} and thereby establish a valuable
tool to find potentially informative partitions, based on the features and
relations of a graph. Furthermore, as we will demonstrate later on, one can
utilize intersection partitions in order to compute similarity measures between
different partitions. We will also make use of intersection partitions in
Sec.~\ref{sec:Imposing Traditional Graph Representations} in order to derive a
tensor-like representation of a multilayer network~\cite{DeDomenico2013}.

To begin with, let us demonstrate what we mean by intersection partitions with a
simple example. Imagine a standard $52$-card deck, partitioned by color on the
one hand (red and black, both comprised of $26$ cards), and by suit on the other
hand (spades, diamonds, hearts and clubs, each comprised of $13$ cards). The
intersection partition of color and suit would then be comprised of $8$
elements: cards that are red and at the same time spades ($0$ cards); red \&
diamonds ($13$ cards); etc. Before showing some examples with regard to
the exemplary graph in Fig.~\ref{fig:example_graph}, let us note down the
different ways of creating intersection partitions of a graph.

We first demonstrate the construction of intersection partitions of $V$. Assume
we are given a set of $K$ [$\leq B(n)$] partitions of $V$, induced by a set of
functions ${}^vp = \lbrace {}^vp^k \;\vline\; k \in I^K \rbrace$, where $I^K =
\lbrace 1, 2, ..., K \rbrace$ is the partition index set. From this set of
available partitions, we choose a collection $g \subseteq I^K$, which is used to
create an intersection partition. We define an element
$V^{\underline{p}}_{\underline{i}}$ of the intersection partition
$V^{\underline{p}}$ by
\begin{align}
&V^{\underline{p}}_{\underline{i}} := \lbrace V_j \;\vline\; j \in \lbrace 1,
2, ..., n \rbrace \land \forall k \in g: {}^vp^k(V_j) = {}^vS^k_{i^k} \rbrace,
\text{ where} \label{eq:ipn1} \\
&\underline{p} = (p^k)_{k \in g}, \underline{i} = (i^k)_{k \in g}, i^k \in
\lbrace 1, 2, ..., n^{p^k} \rbrace, \label{eq:ipn2}
\end{align}
and the intersection partition itself by
\begin{equation}
V^{\underline{p}} := \bigcup_{\underline{i}} V^{\underline{p}}_{\underline{i}}.
\label{eq:ipn3}
\end{equation}
Since $\varnothing \notin V^{\underline{p}}$ by construction, and by showing
that
\begin{align}
&V^{\underline{p}}_{\underline{i}} \cap V^{\underline{p}}_{\underline{j}} =
\varnothing \text{ for all } V^{\underline{p}}_{\underline{i}} \neq
V^{\underline{p}}_{\underline{j}} \in V^{\underline{p}}, \text{ where} \\
&\underline{j} = (j^k)_{k \in g}, j^k \in \lbrace 1, 2, ..., n^{p^k} \rbrace,
\end{align}
we see that $V^{\underline{p}}$ is indeed a partition of $V$. A supernode
$V^{\underline{p}}_{\underline{i}}$ of an intersection partition is comprised
of $n^{\underline{p}, \underline{i}} := |V^{\underline{p}}_{\underline{i}}|$
nodes $V_j$ of $G$ that simultaneously belong to all supernodes ${}^vS^k_{i^k}$
chosen by $g$. The number of supernodes of an intersection partition,
$n^{\underline{p}} := |V^{\underline{p}}|$, is bounded by
$\prod_{k \in g} n^{p^k}$, and every intersection partition constitutes a
refinement of the partitions it has been constructed from, $V^{\underline{p}}
\leq V^{p^k}$ for all $k \in g$. The number of distinct intersection partitions
we can construct from $I^K$ is bounded from above by $I(K) = \sum_{|g|=0}^{K}
\binom{K}{|g|} = |\mathcal{P}(I^K)|$, where $\mathcal{P}(I^K)$ is the power set
of the partition index set, hence $I(K) = 2^K$. Looking at
Fig.~\ref{fig:example_graph}, the intersection partition of the collection of
partitions $g = \lbrace \text{`category'}, \text{`political ideology'}
\rbrace$ would yield $n^{\underline{p}} = 10$ supernodes. The supernodes
comprised of more than $1$ node of $G$ would be: `person' \& `egalitarianism'
(3 nodes); `person' \& `environmentalism' (2 nodes); and `state' \&
`conservatism' (2 nodes).

Defining the corresponding intersection partition $E^{\underline{p}}$ of $E$
into $m^{\underline{p}}: = |E^{\underline{p}}|$ superedges
$E^{\underline{p}}_{\underline{i} \underline{j}}$ by
\begin{equation}
E^{\underline{p}}_{\underline{i} \underline{j}} := \lbrace E_{uv} \;\vline\;
\Phi^e(u,v) \wedge \forall k \in g: {}^vp^k(V_u) = {}^vS^k_{i^k} \wedge \forall
k \in g: {}^vp^k(V_v) = {}^vS^k_{j^k} \rbrace \label{eq:nip}
\end{equation}
and $E^{\underline{p}} := \bigcup_{\underline{i}, \underline{j}}
E^{\underline{p}}_{\underline{i} \underline{j}}$,
it follows that $\underline{p}$ induces a supergraph $G^{\underline{p}} =
(V^{\underline{p}}, E^{\underline{p}})$, in analogy to
Eqs.~(\ref{eq:gp1})-(\ref{eq:gp3}). A superedge
$E^{\underline{p}}_{\underline{i} \underline{j}}$ is comprised of
$m^{\underline{p}, \underline{i}\underline{j}} :=
|E^{\underline{p}}_{\underline{i} \underline{j}}|$ edges $E_{uv}$ originating
from nodes in supernode $V^{\underline{p}}_{\underline{i}}$ and targeting nodes
in supernode $V^{\underline{p}}_{\underline{j}}$.
Figure~\ref{fig:supergraph_lattice}(b) depicts a supergraph, created from
intersecting two different supergraphs.

With regard to partitioning the edges of a graph, however, there are other
options than partitioning by types of relations [see
Eqs.~(\ref{eq:p_edges})-(\ref{eq:ep3})], or by correspondence [see
Eqs.~(\ref{eq:gp1})-(\ref{eq:gp3})]. We now show how to utilize the edges'
relations and the features of their incident nodes in all possible combinations.
For instance, regarding the graph in Fig.~\ref{fig:example_graph}, we might want
to know how many edges originate from nodes with a `political ideology' of
`egalitarianism', or `conservatism', etc. The answer would yield a
total of $m^p = 5$ superedges, originating from: `anarchism' (comprised of 9
edges); `egalitarianism' (5 edges); `conservatism' (2 edges);
`environmentalism' (5 edges); and `neoliberalism' (1 edge). These
superedges, however, could be refined by asking how many of their constituent
edges target nodes of the `category' `bank', or `company' and so forth. We
would then see, for instance, that the edges originating from nodes with a
`political ideology' of `egalitarianism' all target nodes of the
`category' `state'. Refining these superedges even further, we could ask,
how many edges originating from nodes with a `political ideology' of
`egalitarianism' and targeting nodes of the `category' `state' are of the
`kind of payment' `tax', or `bribe', etc. Let us note down all the
combinations formally, to clarify the procedure of partitioning edges.

Assume we are given a set of ${}^vK$ partitions of $V$, induced by ${}^vp =
\lbrace {}^vp^k \;\vline\; k \in {}^vI^{K} \rbrace$, where ${}^vI^K = \lbrace 1,
2, ..., {}^vK \rbrace$ is the partition index set of the nodes. Additionally, we
have a set of ${}^eK$ partitions of $E$, induced by ${}^ep = \lbrace {}^ep^k
\;\vline\; k \in {}^eI^{K} \rbrace$, where ${}^eI^K = \lbrace 1, 2, ..., {}^eK
\rbrace$ is the partition index set of the edges. From these partitions, we
choose three different collections: a \textit{source type collection} $g^s
\subseteq {}^vI^K$, a \textit{target type collection} $g^t \subseteq {}^vI^K$
and a \textit{relation type collection} $g^r \subseteq {}^eI^K$. Then, we denote
a superedge by $E^{\underline{p}}_{\underline{i} \underline{j}, \underline{r}}$,
where
\begin{align}
&\underline{p} = \left( ({}^sp^k)_{k \in g^s}, ({}^tp^k)_{k \in g^t},
({}^rp^k)_{k \in g^r}) \right), \label{eq:ip_edges_first} \\
&\underline{i} = (i^k)_{k \in g^s}, \text{ with } i^k \in \lbrace 1, 2, ...,
n^{p^k} \rbrace, \\
&\underline{j} = (j^k)_{k \in g^t}, \text{ with } j^k \in \lbrace 1, 2, ...,
n^{p^k} \rbrace, \text{ and}\\
&\underline{r} = (r^k)_{k \in g^r}, \text{ with } r^k \in \lbrace 1, 2, ...,
m^{p^k} \rbrace,
\end{align}
and define it by
\begin{align}
&E^{\underline{p}}_{\underline{i} \underline{j}, \underline{r}} :=
\lbrace E_{uv} \;\vline\; \Phi^e(u, v) \land \Phi^v_{g^s}(u) \land
\Phi^v_{g^t}(v) \land \Phi^e_{g^r}(u,v) \rbrace, \text{ where} \\
&\Phi^v_{g^s}(u) : (\forall k \in g^s: {}^vp^k(V_u) = {}^vS^k_{i^k}), \\
&\Phi^v_{g^t}(v) : (\forall k \in g^t: {}^vp^k(V_v) = {}^vS^k_{j^k}),
\text{ and} \\
&\Phi^e_{g^r}(u,v) : (\forall k \in g^r: {}^ep^k(E_{uv}) = {}^eS^k_{r^k}).
\label{eq:ip_edges_last}
\end{align}
The partition $E^{\underline{p}}$ of $E$ is then given by $E^{\underline{p}} :=
\bigcup_{\underline{i}, \underline{j}, \underline{r}}
E^{\underline{p}}_{\underline{i} \underline{j}, \underline{r}}$. Based on these
definitions, we denote the number of superedges by $m^{\underline{p}} =
|E^{\underline{p}}|$, and the number of edges contained in a superedge by
$m^{\underline{p}, \underline{i} \underline{j}, \underline{r}} :=
|E^{\underline{p}}_{\underline{i} \underline{j}, \underline{r}}|$. If all
collections are empty at the same time, $g^x = \varnothing$ for all $x \in
\lbrace s, t, r \rbrace$, it follows that $E^{\underline{p}}_{\underline{i}
\underline{j}, \underline{r}} = E$, which means that the edge set is partitioned
into the trivial partition, comprised of one superedge entailing all edges.
Furthermore, if we choose $g^s = g^t$ and $g^r = \varnothing$, we get the
definition of the corresponding partition, as stated in Eq.~(\ref{eq:nip}).
Expressed formally, the example stated in the above paragraph would hence be
described as follows: we choose the source type collection by $g^s = \lbrace
\text{`political ideology'} \rbrace$, the target type collection by $g^t =
\lbrace \text{`category'} \rbrace$ and the relation type collection by $g^r =
\lbrace \text{`kind of payment'} \rbrace$. The superedge
$E^{\underline{p}}_{\underline{i} \underline{j}, \underline{r}}$ corresponding
to $\underline{i} = (\text{`egalitarianism'})$, $\underline{j} =
(\text{`state'})$ and $\underline{r} = (\text{`tax'})$ would then be
comprised of $m^{\underline{p}, \underline{i} \underline{j}, \underline{r}} = 5$
edges.

Before we turn to the next section, let us make some general remarks regarding
intersection partitions:

i) First of all, it is noteworthy that it only makes sense
to create intersection partitions of antichains, since any chain in $g$, $g^s$,
$g^t$ or $g^r$ can be replaced by the finest element of the respective chain.

ii) When creating intersection partitions we have to be aware of the fact that a
supernode $V^{\underline{p}}_{\underline{i}}$ might be comprised of zero nodes,
$n^{\underline{p}, \underline{i}} = 0$. In this case, we say the supernode
$V^{\underline{p}}_{\underline{i}}$ does not exist. This stands in contrast to
the supernodes $V^{p^k}_i$ of supergraphs $G^{p^k}$, for which $n^{p,i} \geq 1$
for all $i \in \lbrace 1, 2, ..., n^{p^k} \rbrace$, since we chose the functions
${}^vp^k$ to be surjective. This does not pose a problem though, since for a
superedge $E^{\underline{p}}_{\underline{i} \underline{j}}$ with
$m^{\underline{p}, \underline{i}\underline{j}} = 0$, we can still deduce if the
superedge does not exist because at least one of the supernodes does not exist
($n^{\underline{p}, \underline{i}}$ or $n^{\underline{p}, \underline{j}} = 0$),
or because there is in fact no superedge between existing supernodes
($n^{\underline{p}, \underline{i}}$ and $n^{\underline{p}, \underline{j}} \geq
1$).

iii) Finally, we want to refer to Appendix~\ref{sec:Measuring the Similarity of
(Intersection) Partitions}, where we demonstrate how to utilize intersection
partitions in order to compute similarity measures between different
(intersection) partitions. Such measures can be utilized, for instance, to
assess the community structure of time-evolving networks, as the authors
of~\cite{Granell2015} have demonstrated.

\subsection{Redistribution and Allocation of Information on the Lattices}
\label{sec:Redistribution and Allocation of Information on the Lattices}

The last sections were dedicated to constructing partitions, allowing us to
group any desirable subset of nodes and edges into supernodes and superedges,
respectively. Here, we demonstrate that the information of a graph -- expressed
by the features and relations of its constituent nodes and edges -- is not only
conserved under partitioning, but redistributed on the partition lattices,
according to the partition function(s) we choose. This allows us to aggregate
data of any desirable group of nodes or edges. We then demonstrate how to
allocate partition-specific features and relations, which also allows us to
create superedges independently of the edges in $G$.

First, the information contained in a given graph $G = (V, E)$ is conserved when
creating partitions: given a partition $V^{\underline{p}}$ of $V$ induced by
$\underline{p}$ [see Eqs.~(\ref{eq:ipn1})-(\ref{eq:ipn3})], every supernode
$V^{\underline{p}}_{\underline{i}} \in V^{\underline{p}}$ is a subset of the
nodes of $G$, where each node is comprised of a set of features. The complete
set of features contained in supernode $V^{\underline{p}}_{\underline{i}}$ can
then be partitioned by their corresponding types, and therefore expressed as a
collection of sets of features of common type. Hence, a supernode
$V^{\underline{p}}_{\underline{i}}$ -- expressed in terms of its constituent
features -- is given by
\begin{equation}
V^{\underline{p}}_{\underline{i}} = \lbrace \underline{i} \rbrace \cup \lbrace
F^{\underline{p}, T}_{\underline{i}, t} \rbrace_{t \in \lbrace 1, 2, ...,
n^{\underline{p}, \underline{i}}_{\text{types}} \rbrace }, \label{eq:snf}
\end{equation}
where the number of distinct types of features in supernode
$V^{\underline{p}}_{\underline{i}}$ is denoted by $n^{\underline{p},
\underline{i}}_{\text{types}}$, and the number of features of type $t$ by
$n^{\underline{p}, \underline{i}}_{\text{t}} :=
|F^{\underline{p}, T}_{\underline{i}, t}|$. Looking at
Fig.~\ref{fig:example_graph}, the supernode comprised of the nodes with indices
$(2,6,11)$, for instance, has a total of $n^{\underline{p},
\underline{i}}_{\text{types}} = 7$ types of features: `category'
($n^{\underline{p}, \underline{i}}_{1} = 3$: `person': 2 nodes, `company':
1 node); `name' ($n^{\underline{p}, \underline{i}}_{2} = 3$: `Avon
Barksdale': 1 node, `Rust Cohle': 1 node, `Viktor's Gun Shop': 1 node);
`occupation' ($n^{\underline{p}, \underline{i}}_{3} = 2$: `chief executive
officer': 1 node, `federal police officer': 1 node); `age'
($n^{\underline{p}, \underline{i}}_{4} = 3$: `91': 1 node, `61': 1 node,
`141': 1 node); etc. By this example, it becomes clear that we can
easily create frequency distributions of the values of a supernodes' different
types of features.

Analogously, we can express superedges in terms of the relations of their
constituent edges', which we also partition by their corresponding types: given
a partition $E^{\underline{p}}$ of $E$ induced by $\underline{p}$ [see
Eqs.~(\ref{eq:ip_edges_first})-(\ref{eq:ip_edges_last})], a superedge
$E^{\underline{p}}_{\underline{i} \underline{j}, \underline{r}}$ is given by
\begin{equation}
E^{\underline{p}}_{\underline{i}\underline{j}, \underline{r}} = \lbrace
(\underline{i}, \underline{j}, \underline{r}) \rbrace \cup \lbrace
R^{\underline{p}, T}_{\underline{i}\underline{j}, \underline{r}, t}
\rbrace_{t \in \lbrace 1, 2, ..., m^{\underline{p}, \underline{i}\underline{j},
\underline{r}}_{\text{types}} \rbrace }, \label{eq:ser}
\end{equation}
where the number of distinct types of relations in superedge
$E^{\underline{p}}_{\underline{i} \underline{j}, \underline{r}}$ is denoted by
$m^{\underline{p}, \underline{i}\underline{j}, \underline{r}}_{\text{types}}$,
and the number of relations of type $t$ by $m^{\underline{p},
\underline{i}\underline{j}, \underline{r}}_{\text{t}} := |R^{\underline{p},
T}_{\underline{i}\underline{j}, \underline{r}, t}|$. For mathematical details,
we refer to Appendix~\ref{sec:Expressing Supernodes (Superedges) by Features
(Relations)}.

By this representation of supernodes and superedges, we can clearly see that the
information of a graph $G$ is not only conserved under partitioning, but
redistributed according to the partition function(s) we choose. This means that
every supergraph $G^{\underline{p}}$ on the lattice ${}^GL$, and every partition
$E^{\underline{p}}$ on the lattice ${}^EL$, corresponds to a unique
redistribution of the information contained in a graph $G$, and the collection
of all possible redistributions is given by the lattices ${}^GL$ and ${}^EL$.

Second, we show how to allocate partition-specific information on the lattice
${}^GL$. Note that we omit the vector notation of intersection partitions for
the remainder of this section for reasons of notational simplicity. Given a
supergraph $G^p \in {}^GL$, we know that its supernodes are comprised of
features $\lbrace F^{p, T}_{i, t} \rbrace_{t \in \lbrace 1, 2, ...,
n^{p,i}_{\text{types}} \rbrace}$, and its superedges are comprised of relations
$\lbrace R^{p, T}_{ij, t} \rbrace_{t \in \lbrace 1, 2, ..., m^{p,
ij}_{\text{types}} \rbrace}$. Based on these features and relations, we can
compute additional properties (e.g., moments, correlations) by applying some set
of functions on them. For the sake of notational convenience, we write single
functions mapping to sets of new properties:
\begin{align}
&f(\lbrace F^{p, T}_{i,t} \rbrace_{t \in \lbrace 1, 2, ...,
n^{p,i}_{\text{types}} \rbrace}) = \lbrace {}^pF^{p, T}_{i, t} \rbrace_{t \in
\lbrace 1, 2, ..., {}^pn^{p,i}_{\text{types}} \rbrace}, \label{eq:psf} \\
&f(\lbrace R^{p, T}_{ij,t} \rbrace_{t \in \lbrace 1, 2, ...,
m^{p,ij}_{\text{types}} \rbrace}) = \lbrace {}^pR^{p, T}_{ij, t} \rbrace_{t \in
\lbrace 1, 2, ..., {}^pm^{p,ij}_{\text{types}} \rbrace}, \label{eq:psr}
\end{align}
where the additional $p$-index on the upper left corner indicates that these
features and relations are specific to the supergraph $G^p$. Of course, we can
also allocate features to supernodes independently from the features of the
supernodes' constituent nodes. The same goes for the relations of superedges,
even in the case when they are comprised of zero edges (for which $E^{p}_{ij} =
\varnothing$ and therefore also $\lbrace R^{p, T}_{ij,t} \rbrace_{t \in \lbrace
1, 2, ..., m^{p,ij}_{\text{types}} \rbrace} = \varnothing$). We do not, however,
denote these independently allocated features and relations differently to the
computed features and relations in Eqs.~(\ref{eq:psf}) and (\ref{eq:psr}).
Hence, the properties of supernodes and superedges of a supergraph can be
written as
\begin{equation}
V^p_i = \lbrace i \rbrace \cup \lbrace F^{p, T}_{i, t} \rbrace_{t
\in \lbrace 1, 2, ..., n^{p,i}_{\text{types}} \rbrace} \cup \lbrace {}^pF^{p,
T}_{i, t} \rbrace_{t \in \lbrace 1, 2, ..., {}^pn^{p,i}_{\text{types}} \rbrace},
\end{equation}
and
\begin{equation}
E^{p}_{ij} = \lbrace (i, j) \rbrace \cup \lbrace R^{p, T}_{ij, t}
\rbrace_{t \in \lbrace 1, 2, ..., m^{p, ij}_{\text{types}} \rbrace} \cup \lbrace
{}^pR^{p, T}_{ij, t} \rbrace_{t \in \lbrace 1, 2, ...,
{}^pm^{p,ij}_{\text{types}} \rbrace}.
\end{equation}

Of course, the partition-specific features and relations only bear meaning for
the unique element of the lattice $G^{p} \in {}^GL$. Furthermore, they can only
be redistributed on the set of all coarser supergraphs, given by the chains
entailed in $\lbrace G^{p'} \;\vline\; p' \in \lbrace 1, 2, ..., B(n) \rbrace
\land p' > p \rbrace \subseteq {}^GL$ (see the red, dashed lines in
Fig.~\ref{fig:supergraph_lattice}).

\section{Deep Graph Construction} \label{sec:Deep Graph Construction}

The theoretical framework satisfying the objectives stated in the Introduction
is now fully described. Here, we want to roughly describe the general procedure
of constructing a deep graph. For this purpose, we introduce two types of
auxiliary functions: \textit{connectors}, which are functions allowing us to
create (super)edges between (super)nodes, purely based on the properties of the
represented objects; and \textit{selectors}, which are functions allowing us to
select (i.e. filter) (super)edges, based on their respective properties. In
combination, these functions effectively allow us to forge the topology of a
deep graph, which we will exemplify in Sec.~\ref{sec:Application to Global
Precipitation Data}. Furthermore, we demonstrate in this section how our
framework integrates with existing network theory and other data analysis tools,
and finally make some general remarks regarding the identification of
(super)nodes, (super)edges and partitions.

\subsection{Outline of Deep Graph Construction} \label{sec:Outline of Deep Graph
Construction}

Given a set of $n$ objects, the general procedure of constructing a deep graph
can be outlined as follows

\begin{enumerate}
  \item identify each object as a node $V_i$, $i = 1, 2, ..., n$.

  \item assign features to each node $V_i$, $V_i = \lbrace i, F_i^1, F_i^1, ...,
F_i^{f_i} \rbrace$.

  \item define \textit{connectors}
\begin{equation}
m_{ij}(V_i, V_j) := E_{ij} = \lbrace (i, j), R^1_{ij}, R^2_{ij}, ...,
R^{r_{ij}}_{ij} \rbrace,
\end{equation}
where $m_{ij}$ is a function mapping a pair of sets
of features to a set of relations.
Connector functions create ``computable'', or ``external'' relations between
objects. They are typically based on distance or similarity measures of objects,
or some information or physical flow between them. A few examples are the scalar
product of vectors, the distance of objects in a metric space, or correlation
coefficients between variables. Networks solely based on one such measure are
often termed functional networks~\cite{Boers2013,PhysRevLett.97.238103}.

  \item create the set of all possible edges $E'$ by applying the connector
functions on all pairs of nodes.

  \item if there is any a priori knowledge of relations between the objects (as
opposed to the computed relations by connectors), append them to the
corresponding edges. By a priori known relations, we mean any inherent,
internal, physical, trivial or abstract relations, such as flightpaths between
airports, synapses between neurons, social relationships between humans, or
relations of plants to the treatment of medical conditions.

  \item define \textit{selectors}
\begin{equation}
s_{ij}(E_{ij}) := \begin{cases} E_{ij} &\text{if } E_{ij} \text{ satisfies
conditions of } s_{ij} \\ \varnothing & \text {if } E_{ij} \text{ does not
satisfy conditions of } s_{ij} \end{cases},
\end{equation}
where $s_{ij}$ is a function
mapping a set of relations to itself, if the set satisfies the conditions
expressed in the function, or to an empty set otherwise, thereby removing the
corresponding edge from the edge set $E$. Selectors can be simple thresholding
functions (e.g., for some features $F^k_j$ and $F^l_i$: $E_{ij} \mapsto E_{ij}$
if $(F^k_j - F^l_i) \leq T$, else $E_{ij} \mapsto \varnothing$), but they can
also be more complicated and elaborate, involving different types of relations
at the same time.

  \item select $E \subseteq E'$ by applying the selector functions on all edges
$E'$.
\end{enumerate}
The graph is then given by $G = (V, E)$, where the objects' properties are
represented by sets of features $V_i$, and the relational information of pairs
of objects is represented by sets of relations $E_{ij}$.

The next step is to repeat the following procedure for any supergraph $G^p \in
{}^GL$ for which we want to allocate, aggregate or evaluate information. Again,
for notational clarity, we omit vector notation.
\begin{enumerate}

  \item identify a partition $G^p$ of $G$. This partition might be induced by
 the (intersection of) features of the nodes in $G$ (see
 Sec.~\ref{sec:Partitioning Nodes} and Sec.~\ref{sec:Intersection Partitions}),
 or created by any other means, such as manual assignment of supernode labels,
 clustering algorithms, community detection algorithms, or partitioning by the
 connected components of $G$.

  \item compute and allocate partition-specific features to any of the
supernodes \[ V^p_i = \lbrace i \rbrace \cup \lbrace F^{p, T}_{i, t} \rbrace_{t
\in \lbrace 1, 2, ..., n^{p,i}_{\text{types}} \rbrace} \cup \lbrace {}^pF^{p,
T}_{i, t} \rbrace_{t \in \lbrace 1, 2, ..., {}^pn^{p,i}_{\text{types}} \rbrace}.
\]

  \item compute and allocate partition-specific relations to any of the
superedges \[ E^{p}_{ij} = \lbrace (i, j) \rbrace \cup \lbrace R^{p, T}_{ij, t}
\rbrace_{t \in \lbrace 1, 2, ..., m^{p, ij}_{\text{types}} \rbrace} \cup \lbrace
{}^pR^{p, T}_{ij, t} \rbrace_{t \in \lbrace 1, 2, ...,
{}^pm^{p,ij}_{\text{types}} \rbrace}. \]

  \item define \textit{connectors} between supernodes, \[ m_{ij}: V^p \times V^p
\rightarrow E'^p, (V^p_i, V^p_j) \mapsto m_{ij}(V^p_i, V^p_j), \] to further
enrich the relations of the superedges in $G^p$, \[ E^{p}_{ij} = \lbrace (i, j)
\rbrace \cup \lbrace R^{p, T}_{ij, t} \rbrace_{t \in \lbrace 1, 2, ..., m^{p,
ij}_{\text{types}} \rbrace} \cup \lbrace {}^pR^{p, T}_{ij, t} \rbrace_{t \in
\lbrace 1, 2, ..., {}^pm^{p,ij}_{\text{types}} \rbrace} \cup m_{ij}(V^p_i,
V^p_j). \]

  \item define \textit{selectors} on the set of superedges, \[ s_{ij}(E^p_{ij})
  := \begin{cases} E^p_{ij}
&\text{if } E^p_{ij} \text{ satisfies conditions of } s_{ij} \\ \varnothing &
\text{if } E^p_{ij} \text{ does not satisfy conditions of } s_{ij} \end{cases}.
\]

  \item select $E^p \subseteq E'^p$ by applying the selector functions on all
edges $E'^p$.
\end{enumerate}

The supergraph is then represented by $G^p = (V^p, E^p)$. Repeating this
procedure for different elements $G^p \in {}^GL$, we continuously extend the
information contained in ${}^GL$. This information, in turn, can then be
redistributed on the lattice (see Sec.~\ref{sec:Redistribution and Allocation of
Information on the Lattices}, and the red lines in
Fig.~\ref{fig:supergraph_lattice}), and increases the number of possible ways to
create intersection partitions (see Sec.~\ref{sec:Intersection Partitions}, and
the blue lines in Fig.~\ref{fig:supergraph_lattice}).

\subsection{Imposing Traditional Graph Representations}\label{sec:Imposing
Traditional Graph Representations}

Here, we show how to obtain existing network representations, by imposing
certain restrictions on our framework resulting in the multilayer network (MLN)
representation, as defined by Kivel\"a et al~\cite{kivela2014multilayer}. We
chose to demonstrate only the attainment of the MLN representation for two
reasons. First, because it is -- to the best of our knowledge -- the most
general framework of network representation today, and second, because it allows
us refer to the extensive work done by Kivel\"a et
al~\cite{kivela2014multilayer}, Boccaletti et al~\cite{boccaletti2014structure},
and references therein. In these papers, the reader can find derivations of many
additionally constrained network representations down to the level of ordinary
graphs~\cite{Bollobas1998}, as well as a compendium of network tools, models and
concepts to analyze networks. Therefore, the derivation of the MLN
representation -- in conjunction with the work done in these papers -- allows us
to exploit the already existing tool set of network theory.

For readers unfamiliar with MLNs, we provide a summary in
Appendix~\ref{sec:Measuring the Similarity of (Intersection) Partitions}.
Without loss of generality, we assume a MLN $M = (V_M, E_M, V^N, \bm{L})$ with
$|V^N| =: N$ nodes and $|V_M| =: n \leq |V^N| \cdot \prod^{d}_{a=1}|L_a|$
node-layers. First, we have to restrict ourselves to the representation of a
single element of the partition lattice of a deep graph, $G^{p} \in {}^GL$. Let
us assume that this element is the finest element of ${}^GL$ w.l.o.g., $G = (V,
E)$. Then, there are two choices of $G$, resulting in distinct representations
of $M$. We can place the additional information attributed to the layered
structure of $M$ either in the nodes of $G$, or in the edges of $G$. The latter
case is described in Appendix~\ref{sec:Discussion of Multilayer Networks}. The
former case, which is the favourable representation of $M$, is described in the
following.

\begin{figure*}
\includegraphics[width=1\linewidth]{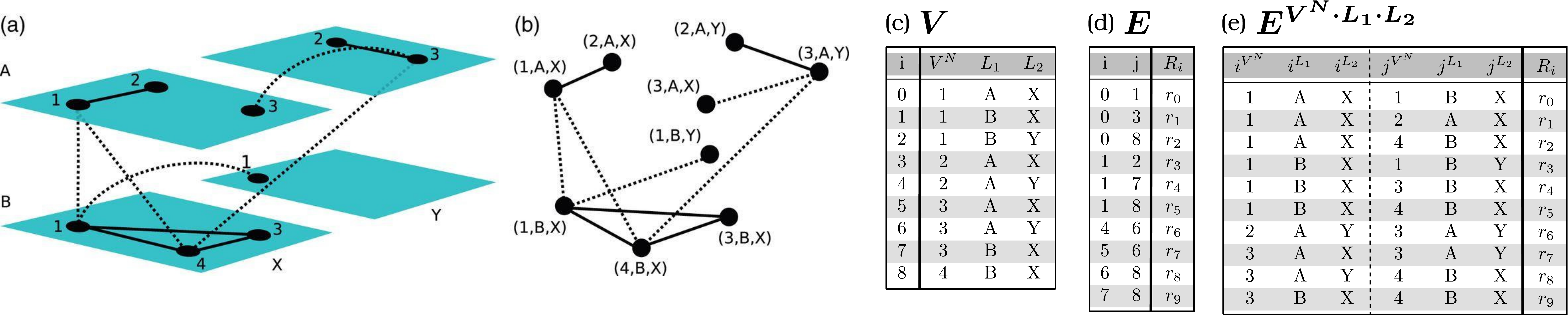}
\caption{\label{fig:mln} \textsf{An exemplary multilayer network (MLN) and its
representation by our framework.} \textbf{(a)} An exemplary MLN, $M = (V_M, E_M,
V^N, \bm{L})$, consisting of four nodes, $V^N = \lbrace 1, 2, 3, 4 \rbrace$, and
two aspects, $\bm{L} = \lbrace L_1, L_2 \rbrace$, where $L_1 = \lbrace A, B
\rbrace$ and $L_2 = \lbrace X, Y \rbrace$. It has a total of 9 node-layers, $V_M
= \lbrace (1,A,X), (1,B,X), (1,B,Y), (2,A,X), (2,A,Y), (3,A,X), (3,A,Y),
(3,B,X), (4,B,X) \rbrace$, connected pair-wise by a total of 10 edges, $E_M
\subset V_M \times V_M$. For notational brevity, we consider the edges to be
directed (with randomly chosen directions).
\textbf{(b)} The same MLN as in (a), depicted by its underlying `supra-graph'
representation, $G_M = (V_M, E_M)$.
\textbf{(c)} The nodes $V_i \in V$ of the graph $G = (V, E)$, representing the
MLN described in (a). $G$ has a total of 9 nodes (corresponding to the MLN's
node-layers), whose indices are given by the left column. The top row indicates
the nodes' \textit{types} of features, which correspond one-to-one to the MLN's
node indices and its aspects.
\textbf{(d)} The edges $E_{ij} \in E$ of the graph $G = (V, E)$, representing
the MLN described in (a). $G$ has a total of 10 edges, corresponding to the
edges $E_M$ of $M$. The first two columns indicate the indices of edges $E_{ij}$
from node $V_i$ to node $V_j$. The (complex- or real-valued) relations of the
edges are denoted by $r_i$ and their corresponding \textit{types} by $R_i$
(which are condensed into one column, for reasons of space).
\textbf{(e)} A tensor-like representation of the edges of the MLN described in
(a). It is derived from the graph $G$ [see (c) and (d)], by constructing the
intersection partition of all its \textit{types} of features, resulting in the
supergraph $G^{V^N \cdot L_1 \cdot L_2} = (V^{V^N \cdot L_1 \cdot L_2}, E^{V^N
\cdot L_1 \cdot L_2})$. The supergraph's edges $E^{V^N \cdot L_1 \cdot
L_2}_{i^{V^N} \cdot i^{L_1} \cdot i^{L_2}, j^{V^N} \cdot j^{L_1} \cdot j^{L_2}}
\in E^{V^N \cdot L_1 \cdot L_2}$ are indexed like a tensor, as apparent from the
table. See Appendix~\ref{sec:Discussion of Multilayer Networks} for mathematical
details. Figure~\ref{fig:mln}(a) and (b) are reproduced with permission from
Journal of Complex Networks 2, 203 - 271 (2014). Copyright 2013 Oxford
University Press - Journals.}
\end{figure*}

We identify each node $V_i \in V = \lbrace V_1, V_2, ..., V_n \rbrace$ with a
node-layer $V_{M,i} \in V_M$, such that
\begin{equation}
V_i = \lbrace V^N_i, L_{1,i}, L_{2,i}, ..., L_{d,i} \rbrace
\mathrel{\widehat{=}} V_{M, i} \in V^N \times L_1 \times L_2 \times \cdot \cdot
\cdot \times L_d, \label{eq:mln1}
\end{equation}
where $V^N_i \in V^N$ and $L_{a,i} \in L_a$ for all $a \in \lbrace 1, 2, ..., d
\rbrace$. This means that every node $V_i$ of $G$ has one feature corresponding
to the index of a node $V^N_i \in V^N$ and $d$ features corresponding to
elementary layers of the aspects $L_a \in \bm{L}$. An edge $E_{ij} \in E' =
\lbrace E_{11}, E_{12}, ..., E_{nn} \rbrace$ is given by
\begin{equation}
E_{ij} =
\begin{cases}
\lbrace w \left( (V_{M,i},V_{M,j}) \right) \rbrace & \text{if }
(V_{M,i},V_{M,j}) \in E_M \\
\varnothing & \text{if } (V_{M,i},V_{M,j}) \notin E_M \end{cases}.
\label{eq:mln2}
\end{equation}
Therefore, the edge set $E$ corresponding to $E_M$ is given by $E = \lbrace
E_{ij} \;\vline\; i,j \in \lbrace 1, 2, ..., n \rbrace \land E_{ij} \neq
\varnothing \rbrace$. Every edge $E_{ij} \in E$ has exactly one relation, whose
type is determined by the tuple of features $(\lbrace L_{a,i}
\rbrace^{d}_{a=1},\lbrace L_{a,j} \rbrace^{d}_{a=1})$ of the adjacent nodes
$V_i$ and $V_j$. The derived representation $G = (V, E)$ corresponds one to one
to the `supra-graph' representation of a MLN, given by the tuple $(V_M, E_M)$.
Figure~\ref{fig:mln} shows an examplary MLN, side by side with its
representation derived here and a tensor-like representation we derive in
Appendix~\ref{sec:Discussion of Multilayer Networks}.

In Appendix~\ref{sec:Discussion of Multilayer Networks}, we demonstrate how the
subset of the partition lattice ${}^GL$ of $G \mathrel{\widehat{=}} M$ induced
by the types of features of its constituent nodes corresponds to different
representations of a MLN, including the above mentioned tensor-like
representation~\cite{DeDomenico2013}. There, we also discuss the constraints
imposed on our framework in order to obtain the MLN representation, and how our
representation solves the issues encountered in the representation of MLNs.

\subsection{Integration with other Data Analysis Tools}\label{sec:Integration
with other Data Analysis Tools}

As demonstrated above, the (super)nodes and (super)edges of a (super)graph
$G^{\underline{p}} \in {}^GL$ are nothing less than collections of sets of
mathematical objects,
\begin{align}
&V^{\underline{p}}_{\underline{i}} = \lbrace \underline{i} \rbrace \cup \lbrace
F^{\underline{p}, T}_{\underline{i}, t} \rbrace_{t \in \lbrace 1, 2, ...,
n^{\underline{p},\underline{i}}_{\text{types}} \rbrace} \cup \lbrace
{}^{\underline{p}}F^{\underline{p}, T}_{\underline{i}, t} \rbrace_{t \in
\lbrace 1, 2, ..., {}^{\underline{p}}n^{\underline{p},
\underline{i}}_{\text{types}} \rbrace}, \text{ and} \\
&E^{\underline{p}}_{\underline{i}\underline{j}} = \lbrace (\underline{i},
\underline{j}) \rbrace \cup \lbrace R^{\underline{p}, T}_{\underline{i}
\underline{j}, t} \rbrace_{t \in \lbrace 1, 2, ..., m^{\underline{p},
\underline{i}\underline{j}}_{\text{types}} \rbrace} \cup \lbrace
{}^{\underline{p}}R^{\underline{p}, T}_{\underline{i}\underline{j}, t}
\rbrace_{t \in \lbrace 1, 2, ..., {}^{\underline{p}}m^{\underline{p},
\underline{i}\underline{j}}_{\text{types}} \rbrace} \cup m_{\underline{i}
\underline{j}}(V^{\underline{p}}_{\underline{i}},
V^{\underline{p}}_{\underline{j}}),
\end{align}
just like the superedges of the partitions $E^{\underline{p}} \in {}^EL$ [see
Eq.~(\ref{eq:ser})]
\begin{equation}
E^{\underline{p}}_{\underline{i}\underline{j}, \underline{r}} = \lbrace
(\underline{i}, \underline{j}, \underline{r}) \rbrace \cup \lbrace
R^{\underline{p}, T}_{\underline{i}\underline{j}, \underline{r}, t} \rbrace_{t
\in \lbrace 1, 2, ..., m^{\underline{p}, \underline{i}\underline{j},
\underline{r}}_{\text{types}} \rbrace }.
\end{equation}
Therefore, there is nothing hindering us from utilizing the tool sets developed
in fields such as multivariate statistics, probability theory, supervised and
unsupervised machine learning, and graph theory, in order to analyze the
properties of (super)nodes and their relations. For instance, we can use
machine learning algorithms to predict missing features of nodes, or to predict
relations between objects. We can use statistical tools to compute properties
such as moments, ranges, covariances and cross-entropies. We can also compute
graph theoretical measures, such as centrality measures (e.g. eigenvector
centralities, betweenness centralities, closeness centralities, degree
centralities), participation coefficients, matching indices or local clustering
coefficients. Furthermore, we can compute similarity or distance measures
through connector functions, and then use appropriate clustering algorithms,
such as stochastic block
models~\cite{aicher2013adapting,Peixoto2012,Peixoto2013,Peixoto2014}, in order
to find informative partitions. All these properties and labels can then be
reassigned to the features and relations of the (super)nodes and (super)edges.

\subsection{Identification of (Super)Nodes, (Super)Edges and
Partitions}\label{sec:Identification of (Super)Nodes, (Super)Edges and
Partitions}

The framework we have laid down offers a good deal of flexibility in mapping
systems onto networks. For that reason, we want to conclude this section by
making a number of general remarks regarding the identification of (super)nodes,
(super)edges, their respective properties and partitions.

i) First of all, recall that the nodes of a graph represent arbitrary objects.
There are no restrictions of what constitutes an object, so a node might
represent literally anything that comes to mind. On top of that, the features of
a node themselves can be arbitrary objects. This means, however, that the
features of a node might themselves be identified as nodes, and vice versa. With
regard to the exemplary graph in Fig.~\ref{fig:example_graph}, for instance, the
nodes with indices $8$ and $9$ (each representing a `state') might just as
well have been identified as features (of type `lives in') of the nodes
representing persons (indices $0$-$6$). Yet, we identified them as nodes
connected by edges (with the type of relation `lives in') to the nodes
$0$-$6$. There are, of course, no general rules of what to identify as features,
and what as nodes. This choice depends mainly on the context.

ii) A similar situation arises when dealing with variables $X = \lbrace x_i
\;\vline\; i \in \lbrace 1, 2, ..., n \rbrace \rbrace$. Imagine, for instance, a
set of variables, each representing a time-series of measurements (e.g. of the
channels in an EEG measurement). Then, each variable can be identified as a
node, whose feature is the variable itself. But we could also identify each
single value assumed by the variables as a node, and include features indicating
the variables (supernodes) each node belongs to. Similar to the identification
of the node-layers (as opposed to the nodes) of a MLN as the nodes of a deep
graph (see Appendix~\ref{sec:Expressing Supernodes (Superedges) by Features
(Relations)}), the latter choice is more flexible, and actually contains the
former choice as supernodes. By identifying each value of a time-series as a
node, for instance, we can create additional supernode labels corresponding to
discretizations of either axis (time or values), such as a discretization into a
certain number of quantiles. Such a concept has been used
in~\cite{campanharo2011duality} to create a map from a time series to a network
with an approximate inverse operation. Within our framework, a bijection between
a variable $X$ and the nodes of a graph $G$ is trivially given by
\begin{align}
&m_{b}: X \leftrightarrow V,  x_i \mapsto m_b(x_i) := V_i = \lbrace i, x_i
\rbrace, \\
&m_b^{-1}(m_b[X]) = X.
\end{align}
Similarly, we can map multidimensional objects (or observations, in machine
learning parlance)
\begin{equation}
X = \lbrace \underline{x}_i = (x^j_i)_{j \in \lbrace 1, 2,..., p \rbrace} \in
\mathbb{R}^p \;\vline\; i \in \lbrace 1, 2, ..., n \rbrace \rbrace
\label{eq:X1}
\end{equation}
to the nodes of a graph $G = (V, E)$, by a function
\begin{align}
&m_{b}: X \leftrightarrow V,  \underline{x}_i \mapsto m_b(\underline{x}_i) :=
V_i = \lbrace i, \underline{x}_i \rbrace, \label{eq:X2}\\
&m_b^{-1}(m_b[X]) = X. \label{X3}
\end{align}
This allows us, for instance, to create edges between objects containing the
derivatives of each pair of variables, $m(V_i, V_j) := E_{ij} = \lbrace
\frac{x^k_j - x^k_i}{x^l_j - x^l_i} \rbrace_{k \neq l \in \lbrace 1, 2, ..., p
\rbrace}$.

iii) It is also straightforward to represent and analyze recurrence
networks~\cite{marwan2008} by deep graphs. Given a $p$-dimensional phase
space and a (discretized) phase-space trajectory represented by a temporal
sequence of $p$-dimensional vectors $\underline{x}_i \in X$ [see
Eq.~\ref{eq:X1}], we first map each point $\underline{x}_i$ of the trajectory to
a node $V_i$ as described in Eq.~\ref{eq:X2}. Then, we create edges between
these nodes, based on some metric on the given phase space (e.g., the euclidian
distance), $m(V_i, V_j) := E_{ij} = \lbrace \| \underline{x}_j - \underline{x}_i
\| \rbrace$. Finally, we define a \textit{selector} $s(E_{ij})$, $s(E_{ij})
\mapsto E_{ij}$ if $\|\underline{x}_j - \underline{x}_i \| < \varepsilon$, else
$s(E_{ij}) \mapsto \varnothing$, leaving only edges between nodes with a
distance smaller than $\varepsilon$, indicating the recurrence of a state in
phase space. The recurrence network is then given by $G = (V, E)$. This approach
can be generalized to cross and joint recurrence networks~\cite{Marwan2002}, by
mapping a collection of phase space trajectories to the nodes of a graph (where
to each node an additional feature is prescribed, indicating the
trajectory it belongs to), and defining \textit{connectors} and
\textit{selectors} accordingly.

iv) As a general rule of thumb, any divisible or separable entity of a system
to be mapped to a deep graph should be divided into separate nodes, and their
membership to the corresponding entity indicated by supernode labels.

v) A convenient manner of representing the time evolution of a network, for
instance, is to take a graph (such as the one illustrated in
Fig.~\ref{fig:example_graph}), and prescribe to every node (edge) a feature
(relation) of the type `time'. Then, one simply copies the nodes and edges of
the graph, indicates their point in time, and adjusts their features and
relations according to whatever properties of the graph have changed over time.
The deep graph incorporating the time evolution of the network is then given by
joining all the copies of nodes and edges that we created into one graph.

vi) In terms of detecting partitions and identifying supernodes, we can also
exploit the topological structure of a graph. In this respect, the auxiliary
\textit{connector} and \textit{selector} functions introduced above constitute a
helpful tool. Given a set of objects, the application of connectors and
selectors allows us to effectively forge the topology of a (super)graph
according to the research question at hand. This is particularly useful for
spatially, temporally or spatio-temporally embedded systems, where we can define
a metric space in which we place the objects of interest. Thereby, for instance,
we may track objects in space over time by connecting them whenever they are
close according to the metric, and then identify the connected components as the
trajectories of the objects. Or, as we will demonstrate in
Sec.~\ref{sec:Application to Global Precipitation Data}, we can use graph
forging as a clustering scheme inducing a partition of the objects, and then
define similarity measures on the induced subgraphs to detect recurrences of
patterns.

vii) Finally, we want to emphasize that the identification of supernodes and
superedges also constitutes a convenient manner of querying a deep graph, by
allowing us to select any desirable group of nodes and edges, in order to
aggregate their respective properties. Such a query could also involve graph
theoretic objects, such as: in- and out-neighbours of a (super)node; paths;
trees; forests; clusters; components; or communities.

\section{Application to Global Precipitation Data}\label{sec:Application to
Global Precipitation Data}

In this section, we demonstrate an application of our framework to a real world
dataset. The basis of this application is the TRMM 3B42 (V7)
dataset~\cite{huffman2007trmm}, comprised of $N = 46.752 \cdot 1440 \cdot 400 $
precipitation measurements from 1998 to 2014, on a spatial resolution of
$0.25^{\circ} \times 0.25^{\circ}$ and a temporal resolution of three hours.
Each data point consists of the time of the measurement $t_i$, the geographical
location, given by a tuple of coordinates $(lon_i, lat_i)$, and the average
precipitation rate of a 3-hour time window $r_i$.

The difference of our approach to previous network-based studies of this dataset
is that we do not create a synchronization-based functional network from the
time-series of precipitation measurements corresponding to the different
geographical locations, as e.g. in~\cite{Stolbova2014,Boers2013,Boers2014SE}.
Instead, we are interested in local formations of spatio-temporal clusters of
extreme precipitation events. For that matter, we first use our framework to
identify such clusters and track their temporal evolution in
Sec.~\ref{sec:Partitioning into Spatio-Temporal Clusters}. Thereafter, we
partition the resulting spatio-temporal clusters into families according to
their spatial overlap in Sec.~\ref{sec:Partitioning into Families of Clusters}.
Finally, climatological interpretations of two exemplary propagation patterns
over the South American continent are provided in Sec.~\ref{sec:Families of
Extreme Rainfall Clusters over South America}

\subsection{Preprocessing of the Data and Identification of the Nodes}

We are only interested in extreme precipitation events, and therefore only
consider 3-hourly measurements above the $90$th percentile of so-called
\textit{wet times} (defined as data points with rainfall rates $r \geq 0.1
\frac{mm}{h}$). The $90$th percentile is chosen in agreement with the definition
of extreme precipitation events in the IPCC report~\cite{Field2012}. These $n
\approx 2.16 \cdot 10^8$ extreme events serve as the data basis for the
following construction of a deep graph. We identify each of the $n$ data points
as a node $V_i$ of the Graph $G = (V, E)$, with $V = \lbrace V_i \;\vline\; i
\in \lbrace 1, 2, ..., n \rbrace \rbrace$. Next, we assign features $F^j_i$ to
the nodes $V_i$ by processing the information given by the dataset as follows.

We enumerate the given longitude, latitude and time coordinates, in order to
associate every node with discrete space-time coordinates, $(lon_i, lat_i, t_i)
\leftrightarrow (x_i, y_i, t_i) =: \underline{x}_i$. By this association, we are
embedding the nodes into a 3-dimensional grid-cell geometry, which we will use
below to identify spatio-temporal clusters. Furthermore, to each tuple $(lon_i,
lat_i)$ we assign a \textit{geographical label}, $(lon_i, lat_i) \leftrightarrow
L_i$, such that nodes with the same geographical location share the same label.
This will enable us to measure spatial overlaps of spatio-temporal clusters
later on. We also compute the surface area $a_i$ and the volume of water
precipitated $v_i$ for each node. Hence, at this stage, every node has a total
of six features, $V_i = \lbrace L_i, \underline{x}_i, a_i, r_i, v_i \rbrace$, as
summarized in Tab.~\ref{tab:g_tables}(a).
\begin{table}
\includegraphics[width=.55\linewidth]{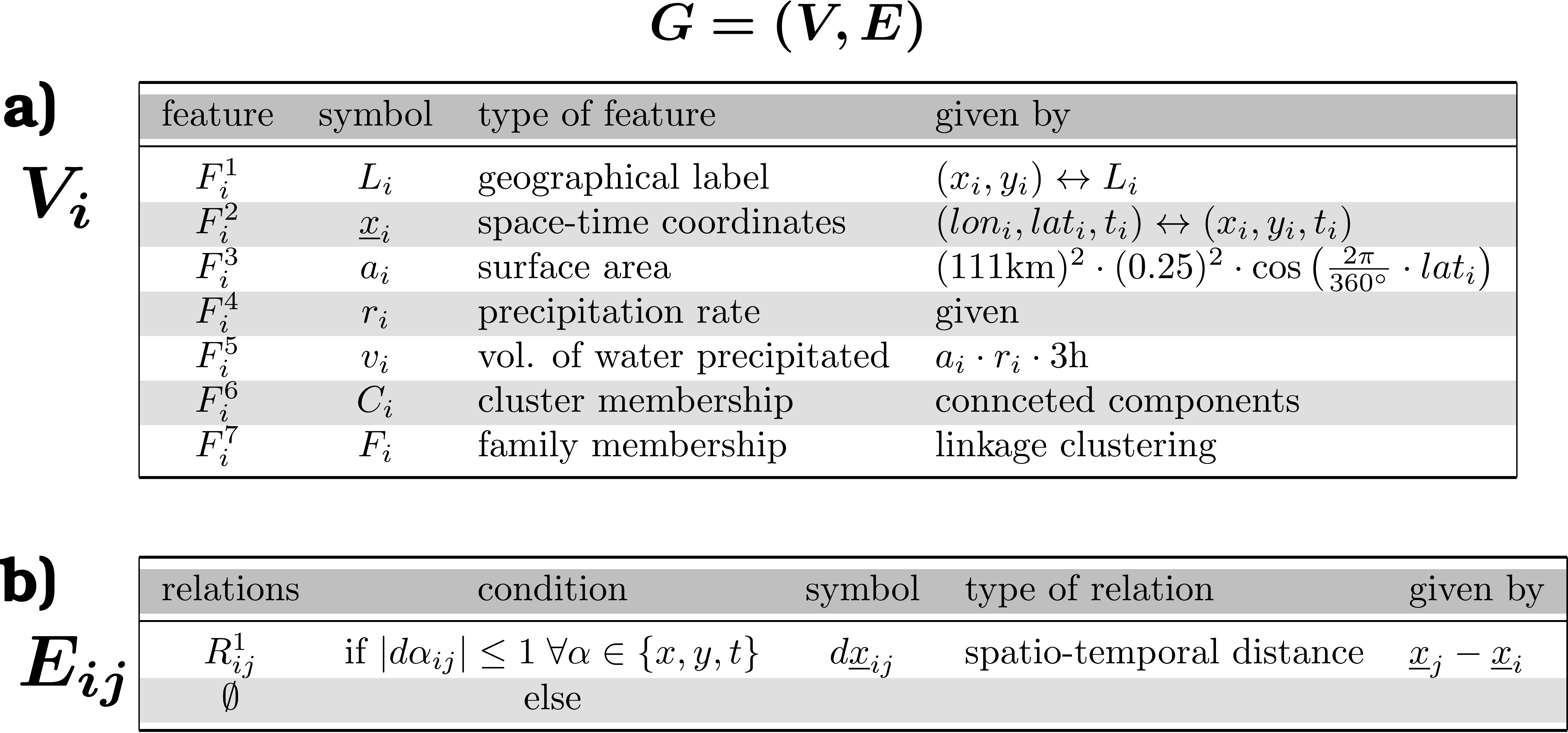}
\caption{\label{tab:g_tables}\textsf{The features and relations of the graph
$G = (V, E)$.} \textbf{(a)} The features of the nodes $V_i$, representing
extreme precipitation measurements. The type of feature `cluster
membership' is introduced in Sec.~\ref{sec:Partitioning into Spatio-Temporal
Clusters} and the type of feature `family membership' in
Sec.~\ref{sec:Partitioning into Families of Clusters}. \textbf{(b)} The
relation of the edges $E_{ij}$, representing the spatio-temporal distance
between precipitation measurements. An edge only exists, if the condition
stated in the table is fulfilled.}
\end{table}

\subsection{Partitioning into Spatio-Temporal Clusters} \label{sec:Partitioning
into Spatio-Temporal Clusters}

As we have assigned the same types of features to all nodes, we can define a
single connector that we apply to all pairs of nodes,
\begin{equation}
m(V_i, V_j) := E_{ij} =
\lbrace (\underline{x}_j - \underline{x}_i) \rbrace =: \lbrace
d\underline{x}_{ij} \rbrace.
\end{equation}
The set of all edges is therefore given by $E' =
\lbrace E_{ij} \;\vline\; i,j \in \lbrace 1, 2, ..., n \rbrace \rbrace$, where
each of the $|E'| \approx 4.69 \cdot 10^{16}$ elements corresponds to a discrete
distance vector of a pair of measurements. The edges of $G$ will be utilized to
detect spatio-temporal clusters in the data, or in more technical terms: to
partition the set of all nodes into subsets of connected grid points. One can
imagine the nodes to be elements of a 3 dimensional grid box, where we allow
every node to have 26 possible neighbours (8 neighbours in the time slice of the
measurement, $t_i$, and 9 neighbours in each the time slice $t_i - 1$ and $t_i +
1$). We can compute the clusters by identifying them as the connected components
of the graph $G = (V, E)$, where $E$ is given by applying the selector
\begin{equation}
s(E_{ij}) := \begin{cases} E_{ij} & \text{if } |d\alpha_{ij}| \leq 1 \forall
\alpha \in \lbrace x, y, t \rbrace \land i \neq j \\ \varnothing & \text{else }
\end{cases}
\end{equation}
on all edges, such that $E = \lbrace E_{ij} \;\vline\; i,j \in
\lbrace 1, 2, ..., n \rbrace \land E_{ij} \neq \varnothing \rbrace$ leaves only
$m = |E| \approx 9.16 \cdot 10^8$ edges between nodes that are neighbours on the
grid.

Identifying the connected components of $G$ results in a labelling of the nodes
according to their respective cluster membership. We find a total of $n^C
\approx 1.42 \cdot 10^7$ spatio-temporal clusters, and transfer their labels as
features to the nodes of $G$, $V_i = \lbrace L_i, \underline{x}_i, a_i, r_i,
v_i, C_i \rbrace$, where $C_i$ indicates to which cluster a node $V_i$ belongs
to. We denote the corresponding partition function by $p^C$, hence $p^C(V_i) =
C_i$. This labelling induces a partition of the graph $G = (V, E)$ into $n^C$
spatio-temporal clusters $V^C_i$ of the supergraph $G^C = (V^C, E^C)$, with
$V^C_i = \lbrace V_j \;\vline\; j \in \lbrace 1,2,...,n \rbrace \land p^C(V_j) =
C_i \rbrace$.

Next, we compute partition-specific features ${}^CF^j_i$ to assign to the
supernodes $V^C_i$, based on the features of the nodes $V_i \in V$. These
features and their calculation are summarized in Tab.~\ref{tab:cg_tables}(a).
\begin{table}
\includegraphics[width=.5\linewidth]{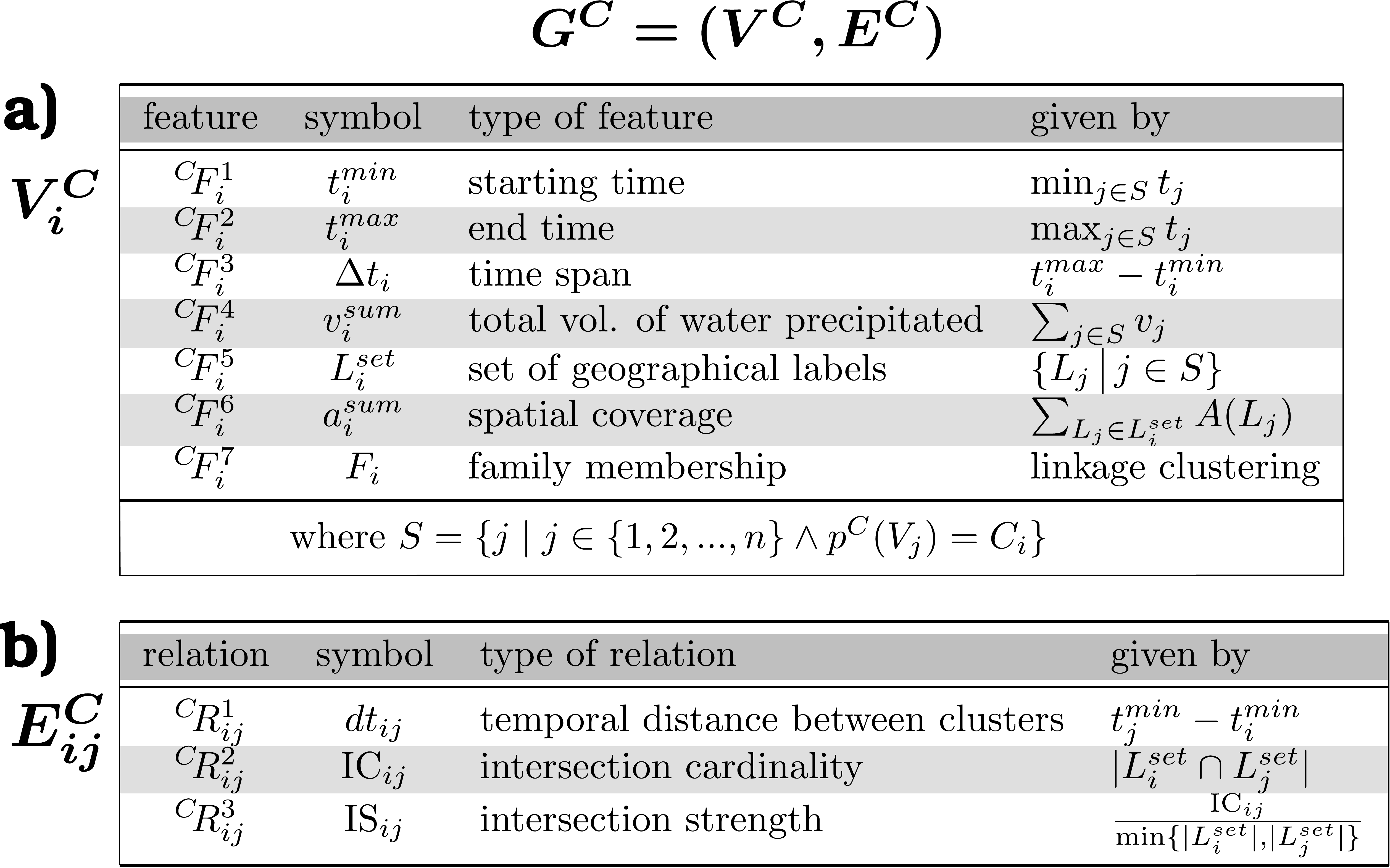}
\caption{\label{tab:cg_tables}\textsf{The features and relations of the
supergraph $G^C = (V^C, E^C)$.} \textbf{(a)} The features of the supernodes
$V^C_i$, representing spatio-temporal clusters of extreme precipitation
measurements. To compute the spatial coverage of a cluster, we map each
geographical grid cell to its surface area, $L_i \mapsto A(L_i)$ (see also the
type of feature `surface area' in Tab.~\ref{tab:g_tables}). The type of
feature `family membership' is introduced in Sec.~\ref{sec:Partitioning
into Families of Clusters}. \textbf{(b)} The relations of the superedges
$E^C_{ij}$.}
\end{table}

\subsection{Partitioning into Families of Clusters} \label{sec:Partitioning into
Families of Clusters}

We now create superedges between the spatio-temporal clusters, in order to find
families of clusters that have a strong regional overlap. Applying the following
partition-specific connector function will provide the information necessary for
this task,
\begin{equation}
m(V^C_i, V^C_j) := E^C_{ij} =  \lbrace dt_{ij}, \text{IC}_{ij},
\text{IS}_{ij} \rbrace,
\end{equation}
where $dt_{ij} = t^{min}_j - t^{min}_i$ is the
temporal distance between a pair of clusters, $\text{IC}_{ij} = |L^{set}_i \cap
L^{set}_j|$ is the intersection cardinality, which is the number of coinciding
geographical grid cells, and $\text{IS}_{ij} = \frac{\text{IC}_{ij}}{\min
\lbrace |L^{set}_i|, |L^{set}_j| \rbrace } \in [0,1]$ is the intersection
strength, a measure for the spatial overlap of a pair of spatio-temporal
clusters. These properties are also summarized in Tab.~\ref{tab:cg_tables}(b).

Based on the above measure of spatial overlap between clusters, we now perform
an agglomerative, hierarchical clustering of the spatio-temporal clusters into
regionally coherent families. We restrict ourselves to the largest $n^c =
40.000$ clusters with respect to their type of feature `total vol. of water
precipitated', since we are only interested in the strongest extreme
precipitation clusters in this paper. We use the UPGMA
algorithm~\cite{Sokal1958Statistical} on the distance vector $\underline{d} =
\left( d_{ij} \right)_{i, j \in \lbrace 1, ..., n^c \rbrace, i < j}$, where
$d_{ij} = d(V^C_i, V^C_j) = 1 - \text{IS}_{ij}$, such that we get a total of
$n^F = 50$ families. We transfer their labels to both the supernodes of $G^C$
and the nodes of $G$, hence $V^C_i = \lbrace t^{min}_i, t^{max}_i, \Delta t_i,
v^{sum}_i, L^{set}_i, a^{sum}_i, F_i \rbrace$ and $V_i = \lbrace L_i,
\underline{x}_i, a_i, r_i, v_i, C_i, F_i \rbrace$, where $F_i$ indicates to
which family the node $V_i$ belongs to. We denote the corresponding partition
function by $p^F$, hence $p^F(V_i) = F_i$.

Next, we identify each family of spatio-temporal clusters as a supernode of the
induced supergraph $G^F = (V^F, E^F)$, where $V^F_i = \lbrace V_j \;\vline\; j
\in \lbrace 1,2,...,n \rbrace \land p^F(V_j) = F_i \rbrace$. Note that, if we
were to take the entire set of spatio-temporal clusters, and not just the
strongest $n^c = 40.000$, this partition would be a further coarse-graining of
the partition induced by $p^C$, $G \leq G^C \leq G^F$. Therefore, we can
redistribute the partition-specific information of $G^C$, in order to compute
the features and relations of $G^F$ as stated in Tab.~\ref{tab:fg_tables}(a) and
(b), respectively.
\begin{table}
\includegraphics[width=.6\linewidth]{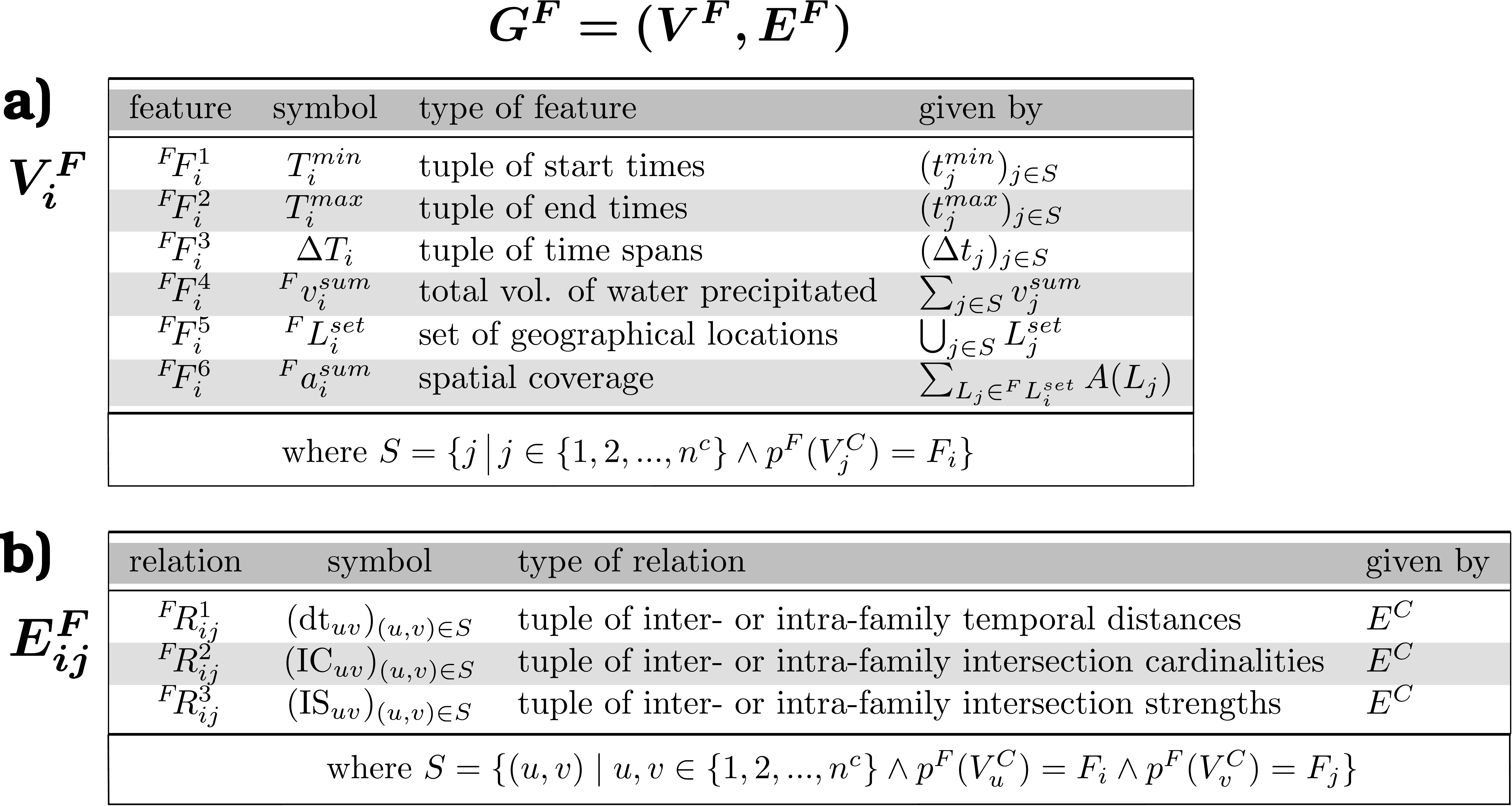}
\caption{\label{tab:fg_tables}\textsf{The features and relations of the
supergraph $G^F = (V^F, E^F)$.} \textbf{(a)} The features of the supernodes
$V^F_i$, representing families of spatio-temporal precipitation clusters. The
first three features are simply the aggregated features of the clusters $V^C_i$.
\textbf{(b)} The relations of the superedges $E^F_{ij}$. They are also just the
unprocessed, aggregated relations between intra-family ($i = j$) and
inter-family ($i \neq j$) clusters.}
\end{table}

We could now compute the temporal inter-cluster intervals of intra-family
clusters, or measure the temporal similarities between families. Indeed, the
information contained in the properties of $G^F$ can easily be mapped onto event
time series. We would only need to identify either $T^{min}_i$ or $T^{max}_i$ as
the time index set $T_i$, and choose the corresponding feature ${}^FF^j_i$ [or
function of features $f({}^FF^1_i, ..., {}^FF^{f_i}_i)$] as the values $v^i$,
\begin{equation}
m_b: V^F \rightarrow X,  V^F_i \mapsto m_b(V^F_i) := X_i = \lbrace v^i_t
\rbrace_{t \in T_i}.
\end{equation}

However, in this paper, we refrain from doing any statistical analysis. Instead,
we demonstrate in the next section how the above created deep graph allows us to
track and visualize the time evolution of extreme precipitation rainfall
clusters.

\subsection{Families of Extreme Rainfall Clusters over South America}
\label{sec:Families of Extreme Rainfall Clusters over South America}

\begin{figure}
\includegraphics[width=.9\linewidth]{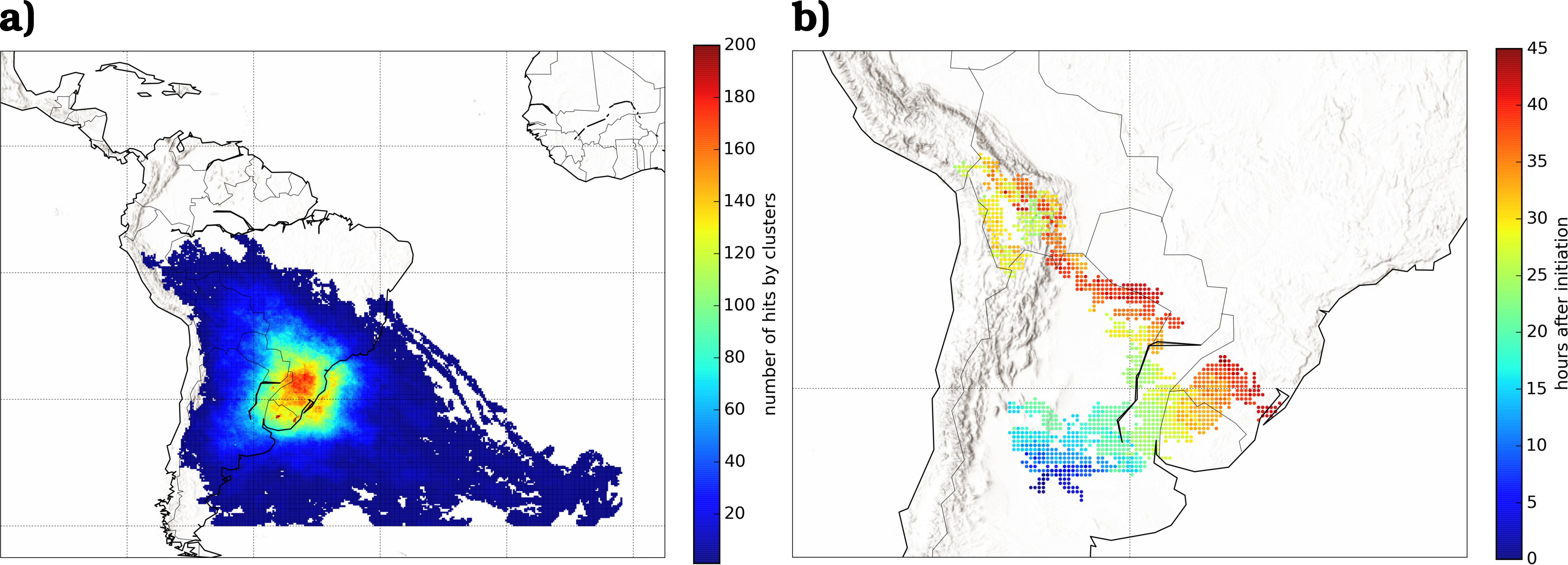}
\caption{\label{fig:subtropic}\textsf{Family of rainfall clusters over
subtropical South America.} \textbf{(a)} The entire family of spatio-temporal
clusters over subtropical South America. The colors indicate how often a given
grid cell $i^L$ is hit by clusters in this family. This number is given by the
number of nodes $n^{FL, i^F i^L}$ in supernode $V^{FL}_{i^F i^L}$ of the
intersection partition $V^{FL}$. Note that the superscript $FL$ indicates that
the supernodes $V^{FL}_{i^F i^L}$ arise from intersecting the partitions given
by the types of features `family membership' $F$ and `geographical label'
$L$. High values over southeastern South America therefore indicate that this
is the core region of this family, where most of its clusters pass by in course
of their lifetime. \textbf{(b)} Exemplary cluster of this family. Each colored
grid cell has received at least one event above the $90$th percentile belonging
to this cluster. The colors indicate the last time (in units of hours) a given
grid cell is hit by the cluster, relative to its initiation on February 6, 2011,
18:00 UTC. The temporal evolution of this cluster therefore shows a concise
propagation pattern from the Argentinean lowlands across Uruguay toward the
eastern slopes of the Central Andes in Bolivia, where the clusters ends on
February 8, 2011, 15:00 UTC. This cluster thus lasted for $\Delta t_i = 45h$,
and the total sum of water it precipitated was $v^{sum}_i = 4.08 \cdot 10^{10}
m^3$, over a total area of $a^{sum}_i = 9.39 \cdot 10^5 km^2$.}
\end{figure}

\begin{figure}
\includegraphics[width=1\linewidth]{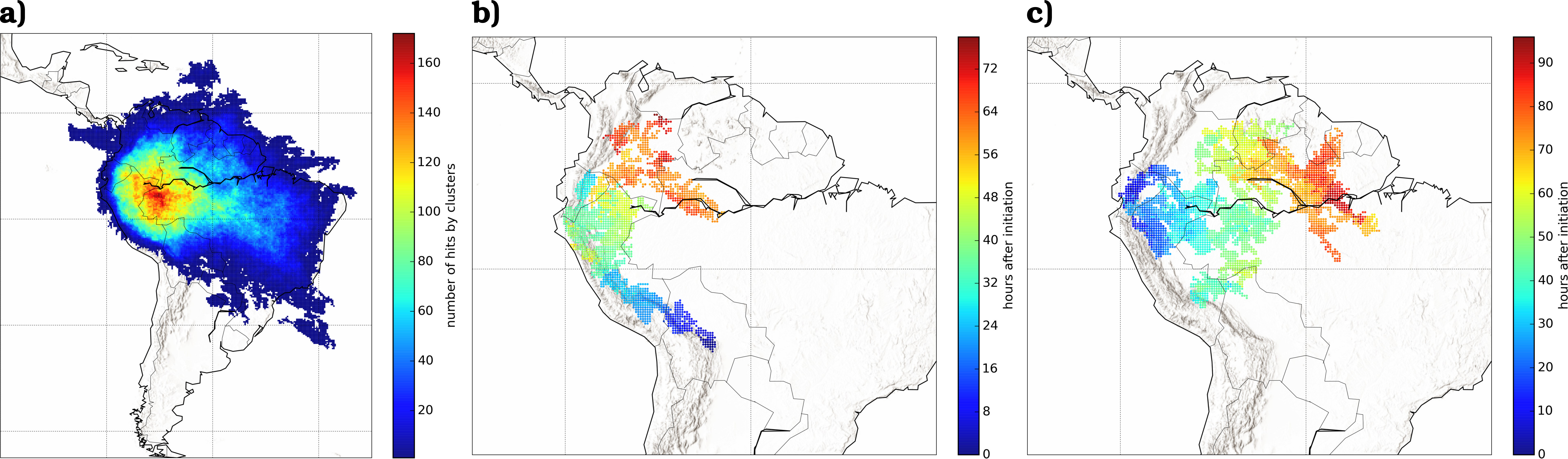}
\caption{\label{fig:tropic}\textsf{Family of rainfall clusters over tropical
South America.} \textbf{(a)} The entire family of spatio-temporal clusters over
tropical South America. The colors indicate how often a given grid cell $i^L$
is hit by clusters in this family, given by $n^{FL, i^F i^L}$ [see the caption
of Fig.~\ref{fig:subtropic}(a)]. High values over the western Amazon therefore
indicate that this the core region of this family, where most of its clusters
pass by in course of their lifetime. \textbf{(b)} First exemplary cluster of
this family. Each colored grid cell has received at least one event above the
$90$th percentile belonging to this cluster. The colors indicate the last time
(in units of hours) a given grid cell is hit by the cluster, relative to its
initiation on November 4, 2002, 9:00 UTC. The temporal evolution of this
cluster therefore shows a concise propagation pattern from central Bolivia
northward, along the eastern slopes of the Andes mountain range, before turning
west in northern Peru. The cluster ends on November 7, 2002, 15:00 UTC over
Colombia and northwestern Brazil, resulting in a total lifetime of
$\Delta t_i = 78h$. The total sum of water precipitated by this cluster is
$v^{sum}_i = 4.90 \cdot 10^{10} m^3$, covering a total area of $a^{sum}_i =
1.53 \cdot 10^6 km^2$. \textbf{(c)} Second exemplary cluster of this family.
It initiated on March 20, 2013, 18:00 UTC, at the eastern slopes of the
northern Peruvian Andes, and thereafter propagated eastward across the entire
Amazon basin, ending on March 24, 2013, 18:00 UTC over northern Brazil. During
its lifetime of $\Delta t_i = 96h$, the total sum of water precipitated by this
cluster is $v^{sum}_i = 1.17 \cdot 10^{11} m^3$, covering a total area of
$a^{sum}_i = 2.61 \cdot 10^6 km^2$.}
\end{figure}

In the following, we restrict ourselves to two families of spatio-temporal
extreme event clusters located over the South American continent. The first
family is confined to the subtropical domain [roughly between $40^{\circ}S$ and
$20^{\circ}S$, see Fig.~\ref{fig:subtropic}(a)], while the second is centered
over the tropical Amazon region [roughly between $10^{\circ}S$ and
$10^{\circ}N$, see~Fig.~\ref{fig:tropic}(a)].

The first family [Fig.~\ref{fig:subtropic}(a)] contains spatio-temporal clusters
of extreme events which are characterized by a concise propagation pattern from
southeastern South America (around $30^{\circ}S$, $60^{\circ}W$) northwestward
to the eastern slopes of the northern Argentinean and Bolivian Andes [see
Fig.~\ref{fig:subtropic}(b) for an example cluster in this family]. These
clusters are remarkable from a meteorological point of view, as their direction
of propagation appears to be against the low-level wind direction in this
region, which is typically from NW to SE~\cite{Vera:2006pi,marengo2012recent}. A
case study based on infrared satellite images~\cite{Anabor2008} analyzes some of
the ``upstream propagating'' clusters in this family in detail. This study,
together with a detailed climatological analysis of these events using the TRMM
3B42 dataset~\cite{Boers2015}, reveals that these spatio-temporal clusters are
in fact comprised of sequences of Mesoscale Convective
Systems~\cite{maddox1980meoscale,durkee2009contribution,durkee2010climatology},
which form successively along the pathway from southeastern South America
towards the Central Andes. The synoptic mechanism explaining this phenomenon is
based on the interplay of cold frontal systems approaching from the South, a
climatological low-pressure system of northwestern Argentina, and low-level
atmospheric moisture flow originating from the tropics~\cite{Boers2015}:
extensive low-pressure systems associated with Rossby wave trains emanating from
the southern Pacific Ocean merge with the low-pressure system over northwestern
Argentina to produce a saddle point of the isobars. Due to the eastward movement
of the Rossby wave train, the configuration of the two low-pressure systems
changes such that the saddle point moves from southeastern South America towards
the Central Andes. The deformation of winds around this saddle point leads to
strong frontogenesis and hence creates favorable conditions for the development
of large-scale organized convection, which explains the observed formation of
several mesoscale convective systems along the pathway this saddle point takes.
Due to the large spatial extents of these rainfall cluster, as well as due the
fact that they propagate into high elevations of the Andean orogen, these
systems impose substantial risks in form of flash-floods and landslides, with
severe consequences for the local populations. Since this pattern is a recurring
feature of the South American Climate system, a complex network approach could
recently be employed to formulate a simple statistical forecast rule, which
predicts more than $60\%$ of extreme rainfall events at the eastern slopes of
the Central Andes~\cite{Boers2014SE}.

The second family [Fig.~\ref{fig:tropic}(a)] we want to show includes
spatio-temporal clusters which exhibit equally concise propagation patterns in
the tropical parts of South America. Similarly to the case described in the
previous paragraph, we find several tropical clusters which propagate in the
opposite direction of the climatological low-level wind fields. Some of these
are initiated at the boundary between tropics and subtropics, move northward
along the eastern slopes of the Peruvian Andes, before turning eastward toward
the Amazonian lowlands [as for example the cluster shown in
Fig.~\ref{fig:tropic}(b)]. Other instances form just east of the northern Andes,
and roughly follow the equator toward the East [as for example the cluster shown
in Fig.~\ref{fig:tropic}(c)]. In view of the above explanations for the first
family, we speculate that similar mechanisms leading to the ``upstream''
propagation of favorable conditions for organized convection are at work in
these cases. However, frontal systems do rarely reach these tropical
latitudes~\cite{Siqueira2005a}, and a saddle point similar to the one described
above is not present in this case. While Amazonian squall lines, which propagate
from the northern Brazilian coast into the continent, have been thoroughly
analyzed~\cite{Tulich2012,Cohen1995}, these organized spatio-temporal clusters
moving northward along the tropical Andes and from West to East across the
Amazon have -- to our knowledge -- not yet been studied in the meteorological
and climatological literature. We therefore propose these particular
spatio-temporal clusters as a promising subject for further research.

\section{Conclusion}\label{sec:Conclusion}

In this paper, we have introduced a collection of definitions resulting in
\textit{deep graphs}, a theoretical framework to describe and analyze
heterogeneous systems across scales, based on network theory. Our framework
unifies existing network representations and generalizes them by fulfilling two
essential objectives: an explicit incorporation of information or data, and a
comprehensive treatment of groups of objects and their relations. The former
objective is implemented by specifying the nodes and edges of a (super)graph as
sets of their respective properties. These properties, which may differ from
node to node and from edge to edge, can be arbitrary mathematical objects.
The second objective is implemented by transferring the mathematical concept of
partition lattices to our graph representation. We have demonstrated how
partitioning the node and edge set of a graph facilitates the means to
aggregate, compute and allocate information on and between arbitrary groups of
nodes. This information can then be stored on the lattices of a graph, allowing
us to express and study properties, relations and interactions on all scales of
the represented system(s).

Based on our representation, we were able to show how deep graphs establish an
interface for common data analysis and modelling tools. This includes
network-based concepts, models and methods, since we derived the different
representations of a multilayer network~\cite{kivela2014multilayer}, which was
the most general network representation to date.

Yet, we have also introduced additional tools to support a comprehensive data
analysis. We have demonstrated how the auxiliary \textit{connector} and
\textit{selector} functions enable us to create and select (super)edges, thereby
allowing us to forge the topology of a deep graph. Intersection partitions not
only allow us to derive a tensor-like representation of a multilayer
network~\cite{DeDomenico2013}, but they also allow us to calculate similarity
measures between (intersection) partitions of a graph and to express elaborate
queries on the information contained in a deep graph.

We have demonstrated some capabilities of our framework by applying it to a
global high-resolution precipitation dataset derived from satellite
measurements. Deep graphs provided a natural and straightforward way to identify
large clusters of extreme precipitation events, track their temporal resolution,
and group the resulting spatio-temporal clusters into families according to
their regional overlap. We have furthermore discussed some climatological
characteristics of two of these families over the South American continent. The
first, which is concentrated over the subtropics, was just recently discovered
using rather complicated methodologies, while the second, which is concentrated
over tropical South America, has to our knowledge not yet been identified and
analyzed in the meteorological literature.

The software package we provide in~\cite{deepgraph} includes all the
capabilities of our representation as described in this paper, and constitutes a
powerful, general-purpose data analysis toolkit. Connector and selector
functions can be defined by the user, which are then combined in order to
efficiently create edges, where the number of CPUs and memory usage can be fully
adjusted.

We hope that our framework initiates attempts to generalize existing network
measures and to develop new measures, particularly in respect of the
heterogeneity of a system's components and their interactions on different
scales. In the context of multilayer networks, generalizations of network
measures have already led to significant new insights, and we expect the same to
become true for deep graphs.

\appendix

\section{Measuring the Similarity of (Intersection)
Partitions}\label{sec:Measuring the Similarity of (Intersection) Partitions}

In this section, we demonstrate how the construction of intersection partitions
provides us with the elements of a so-called confusion matrix (or contingency
table). These are necessary to compute similarity measures between partitions,
such as, e.g.: the Jaccard index~\cite{jaccard1912distribution}; the normalized
mutual information~\cite{strehl2003cluster}; or the normalized variation of
information metric~\cite{meilua2007comparing}. First, we show how to compute the
similarity of two ``normal'' partitions, and then how to compute the similarity
of two intersection partitions.

Assume we are given a graph $G = (V, E)$ comprised of $n$ nodes, and two
partitions of the node set, $V^p = \lbrace V^p_i \;\vline\; i = 1, 2, ..., n^p
\rbrace$ and $V^{p'} = \lbrace V^{p'}_{i'} \;\vline\; i' = 1, 2, ..., n^{p'}
\rbrace$. The number of nodes in supernode $V^p_i$ ($V^{p'}_{i'}$) is then given
by $n^{p,i}$ ($n^{p', i'}$), and the number of nodes in supernode $V^{p \cdot
p'}_{i \cdot i'}$ of the intersection partition $V^{p \cdot p'}$ is given by
$n^{p \cdot p', i \cdot i'}$ [see Eqs.~(\ref{eq:ipn1})-(\ref{eq:ipn3})]. With
these numbers, we can calculate the normalized variation of information metric
by
\begin{equation}
\text{NVI} = \dfrac{-1}{\log n} \sum_{i} \sum_{i'} \dfrac{n^{p \cdot p', i
\cdot i'}}{n} \log \dfrac{(n^{p \cdot p', i \cdot i'})^2}{n^{p,i} n^{p', i'}}.
\label{eq:NVI}
\end{equation}
Analogously, we can compute other similarity measures, such as the Jaccard index
or the normalized mutual information index (see Eqs. (6) and (7)
in~\cite{Granell2015}).

More generally, we can compute the similarity of two intersection partitions.
Assume we are given a graph $G = (V, E)$ comprised of $n$ nodes, and set of $K$
partitions of $V$, induced by a set of functions ${}^vp = \lbrace {}^vp^k
\;\vline\; k \in I^K \rbrace$, where $I^K = \lbrace 1, 2, ..., K \rbrace$ is the
partition index set. From this set of available partitions, we choose two
collections, $g \subseteq I^K$ and $g' \subseteq I^K$, whose corresponding
intersection partitions we want to compare. The number of nodes in supernode
$V^{\underline{p}}_{\underline{i}}$ ($V^{\underline{p}'}_{\underline{i}'}$) is
given by $n^{\underline{p}, \underline{i}}$ ($n^{\underline{p}',
\underline{i}'}$), and the number of nodes in supernode $V^{\underline{p} \cdot
\underline{p}'}_{\underline{i} \cdot \underline{i}'}$ of the intersection
partition $V^{\underline{p} \cdot \underline{p}'}$ is given by $n^{\underline{p}
\cdot \underline{p}', \underline{i} \cdot \underline{i}'}$ (where $\underline{p}
= (p^k)_{k \in g}$, $\underline{i} = (i^k)_{k \in g}$, $\underline{p}' =
(p^k)_{k \in g'}$, $\underline{i}' = (i^k)_{k \in g'}$, and $i^k \in \lbrace 1,
2, ..., n^{p^k} \rbrace$). Using these numbers in Eq.~(\ref{eq:NVI}), we can
compute the similarity of two different intersection partitions.

Equivalently, we can use the numbers $m, m^{\underline{p},
\underline{i}\underline{j}, \underline{r}}, m^{\underline{p}',
\underline{i}'\underline{j}', \underline{r}'}$ and
$m^{\underline{p}\underline{p}',
\underline{i}\underline{i}'\underline{j}\underline{j}',
\underline{r}\underline{r}'}$ (see Tab.~\ref{tab:glossary}) to calculate
similarity measures between (intersection) partitions of the edge set.
Furthermore, we can use a pair of (intersection) partitions of the node set, in
order to compute the similarity of their \textit{corresponding} edge set
partitions.

\section{Expressing Supernodes (Superedges) by Features (Relations)}
\label{sec:Expressing Supernodes (Superedges) by Features (Relations)}

Here, we explicitly demonstrate how the information contained in a given graph
$G = (V, E)$ is conserved when creating partitions, by expressing supernodes and
superedges in terms of features and relations, respectively. Given a partition
$V^{\underline{p}}$ of $V$ induced by $\underline{p}$ [see
Eqs.~(\ref{eq:ipn1})-(\ref{eq:ipn3})], the set of features contained in
supernode $V^{\underline{p}}_{\underline{i}}$ is given by
\begin{equation}
F^{\underline{p}}_{\underline{i}} = \lbrace F^m_j \;\vline\; j \in \lbrace 1, 2,
..., n \rbrace \land m \in \lbrace 1, 2, ..., f_j \rbrace \land \forall k \in g:
{}^vp^k(V_j) = {}^vS^k_{i^k} \rbrace.
\end{equation}
To keep track of a features' original
node index, and to guarantee uniqueness of every single feature, we technically
would have to write $(j, F^m_j)$ for every feature. Yet, for ease of notation,
we refrain from doing so. Next, we map each feature $F^m_j$ in
$F^{\underline{p}}_{\underline{i}}$ onto its respective type,
\begin{equation}
t^{\underline{p}}_{\underline{i}}: F^{\underline{p}}_{\underline{i}} \rightarrow
T^{\underline{p}}_{\underline{i}} = \lbrace 1, 2, ..., n^{\underline{p},
\underline{i}}_{\text{types}} \rbrace,  F^m_j \mapsto
t^{\underline{p}}_{\underline{i}}(F^m_j) := T^{\underline{p}}_{\underline{i}, t}
\in T^{\underline{p}}_{\underline{i}},
\end{equation}
such that
$t^{\underline{p}}_{\underline{i}}(F^l_j) =
t^{\underline{p}}_{\underline{i}}(F^m_k)$ for all pairs of features in
$F^{\underline{p}}_{\underline{i}}$ that share the same type. We denote the
number of distinct types of features in supernode
$V^{\underline{p}}_{\underline{i}}$ by $n^{\underline{p},
\underline{i}}_{\text{types}}$. Note that $0 \leq n^{\underline{p},
\underline{i}}_{\text{types}} \leq |F^{\underline{p}}_{\underline{i}}|$, where
$n^{\underline{p}, \underline{i}}_{\text{types}} = 0$ either because the
supernode $V^{\underline{p}}_{\underline{i}}$ does not exist, $n^{\underline{p},
\underline{i}} = 0$, or because all the nodes it contains have no features,
$n^{\underline{p}, \underline{i}} \geq 1$ and $V_j = \lbrace j \rbrace$ for all
$V_j \in V^{\underline{p}}_{\underline{i}}$. If no pair of nodes in
$V^{\underline{p}}_{\underline{i}}$ shares any type of feature, then
$n^{\underline{p}, \underline{i}}_{\text{types}} =
|F^{\underline{p}}_{\underline{i}}|$. The function
$t^{\underline{p}}_{\underline{i}}$ induces a partition $F^{\underline{p},
T}_{\underline{i}}$ of $F^{\underline{p}}_{\underline{i}}$ into features of
common type $F^{\underline{p}, T}_{\underline{i}, t}$, given by
\begin{equation}
F^{\underline{p}, T}_{\underline{i}, t} = \lbrace F^m_j \;\vline\; j \in \lbrace
1, 2, ..., n \rbrace \land m \in \lbrace 1, 2, ..., f_j \rbrace \land \forall k
\in g: {}^vp^k(V_j) = {}^vS^k_{i^k} \land
t^{\underline{p}}_{\underline{i}}(F^m_j) = T^{\underline{p}}_{\underline{i}, t}
\rbrace,
\end{equation}
and $F^{\underline{p}, T}_{\underline{i}} = \lbrace
F^{\underline{p}, T}_{\underline{i}, t} \;\vline\; t \in \lbrace 1, 2, ...,
n^{\underline{p}, \underline{i}}_{\text{types}} \rbrace \rbrace$. We denote the
number of features of type $t$ in supernode $V^{\underline{p}}_{\underline{i}}$
by $n^{\underline{p}, \underline{i}}_{\text{t}} := |F^{\underline{p},
T}_{\underline{i}, t}|$. Hence, we can express a supernode
$V^{\underline{p}}_{\underline{i}}$ as a set of sets of features of common type
(and its index, to guarantee uniqueness of the supernodes),
\begin{equation}
V^{\underline{p}}_{\underline{i}} = \lbrace \underline{i} \rbrace \cup \lbrace
F^{\underline{p}, T}_{\underline{i}, t} \rbrace_{t \in \lbrace 1, 2, ...,
n^{\underline{p}, \underline{i}}_{\text{types}} \rbrace }.
\end{equation}

Analogously, we can express superedges in terms of their edges' constituent
relations. Given a partition $E^{\underline{p}}$ of $E$ induced by
$\underline{p}$ [see Eqs.~(\ref{eq:ip_edges_first})-(\ref{eq:ip_edges_last})],
the set of relations contained in superedge $E^{\underline{p}}_{\underline{i}
\underline{j}, \underline{r}}$ is given by
\begin{equation}
R^{\underline{p}}_{\underline{i}
\underline{j}, \underline{r}} = \lbrace R^m_{uv} \;\vline\; \Phi^e(u, v) \land
\Phi^v_{g^s}(u) \land \Phi^v_{g^t}(v) \land \Phi^e_{g^r}(u,v) \land m \in
\lbrace 1, 2, ..., r_{uv} \rbrace \rbrace.
\end{equation}
Again, to keep track of a
relations' original indices and to guarantee uniqueness, we technically have to
write $((u,v),R^m_{uv})$ for every relation, which we omit for notational
clarity. Next, we map every relation $R^m_{uv}$ in
$R^{\underline{p}}_{\underline{i} \underline{j}, \underline{r}}$ onto its
respective type,
\begin{equation}
t^{\underline{p}}_{\underline{i}\underline{j},
\underline{r}}: R^{\underline{p}}_{\underline{i}\underline{j}, \underline{r}}
\rightarrow T^{\underline{p}}_{\underline{i}\underline{j}, \underline{r}} =
\lbrace 1, 2, ..., m^{\underline{p}, \underline{i}\underline{j},
\underline{r}}_{\text{types}} \rbrace,  R^m_{uv} \mapsto
t^{\underline{p}}_{\underline{i}\underline{j}, \underline{r}}(R^m_{uv}) :=
T^{\underline{p}}_{\underline{i}\underline{j}, \underline{r}, t} \in
T^{\underline{p}}_{\underline{i}\underline{j}, \underline{r}},
\end{equation}
such that
$t^{\underline{p}}_{\underline{i}\underline{j}, \underline{r}}(R^m_{ij}) =
t^{\underline{p}}_{\underline{i}\underline{j}, \underline{r}}(R^n_{kl})$ for all
pairs of relations in  $R^{\underline{p}}_{\underline{i} \underline{j},
\underline{r}}$ that share the same type. We denote the number of distinct types
of relations in superedge $E^{\underline{p}}_{\underline{i} \underline{j},
\underline{r}}$ by $m^{\underline{p}, \underline{i}\underline{j},
\underline{r}}_{\text{types}}$. Again, $0 \leq m^{\underline{p},
\underline{i}\underline{j}, \underline{r}}_{\text{types}} \leq
|R^{\underline{p}}_{\underline{i}\underline{j}}|$, where $m^{\underline{p},
\underline{i}\underline{j}, \underline{r}}_{\text{types}} =
|R^{\underline{p}}_{\underline{i}\underline{j}}|$ only if no pair of edges in
$E^{\underline{p}}_{\underline{i}\underline{j}, \underline{r}}$ shares any type
of relation. The partition $R^{\underline{p}, T}_{\underline{i}\underline{j},
\underline{r}}$ of $R^{\underline{p}}_{\underline{i}\underline{j},
\underline{r}}$ into relations of common type $R^{\underline{p},
T}_{\underline{i}\underline{j}, \underline{r}, t}$ is therefore induced by the
function $t^{\underline{p}}_{\underline{i}\underline{j}, \underline{r}}$, where
\begin{equation}
R^{\underline{p}, T}_{\underline{i}\underline{j}, \underline{r}, t} = \lbrace
R^m_{uv} \;\vline\; \Phi^e(u, v) \land \Phi^v_{g^s}(u) \land \Phi^v_{g^t}(v)
\land \Phi^e_{g^r}(u,v) \land m \in \lbrace 1, 2, ..., r_{uv} \rbrace \land
t^{\underline{p}}_{\underline{i}\underline{j}, \underline{r}}(R^m_{uv}) =
T^{\underline{p}}_{\underline{i}\underline{j}, \underline{r}, t} \rbrace,
\end{equation}
and
$R^{\underline{p}, T}_{\underline{i}\underline{j}, \underline{r}} = \lbrace
R^{\underline{p}, T}_{\underline{i}\underline{j}, \underline{r}, t} \;\vline\; t
\in \lbrace 1, 2, ..., m^{\underline{p}, \underline{i}\underline{j},
\underline{r}}_{\text{types}} \rbrace \rbrace$. We denote the number of
relations of type $t$ in superedge $E^{\underline{p}}_{\underline{i}
\underline{j}, \underline{r}}$ by $m^{\underline{p}, \underline{i}\underline{j},
\underline{r}}_{\text{t}} := |R^{\underline{p}, T}_{\underline{i}\underline{j},
\underline{r}, t}|$. Therefore, a superedge $E^{\underline{p}}_{\underline{i}
\underline{j}, \underline{r}}$ can be expressed as a set of sets of relations of
common type (and its index, to guarantee uniqueness of the superedges),
\begin{equation}
E^{\underline{p}}_{\underline{i}\underline{j}, \underline{r}} = \lbrace
(\underline{i}, \underline{j}, \underline{r}) \rbrace \cup \lbrace
R^{\underline{p}, T}_{\underline{i}\underline{j}, \underline{r}, t} \rbrace_{t
\in \lbrace 1, 2, ..., m^{\underline{p}, \underline{i}\underline{j},
\underline{r}}_{\text{types}} \rbrace }.
\end{equation}

\section{Summary of the Multilayer Network Representation} \label{sec:Summary of
the Multilayer Network Representation}

In the following, we summarize the representations of a multilayer network
(MLN), as defined by Kivel\"a et al~\cite{kivela2014multilayer}, and refer to
the original paper for a more detailed description. A multilayer network (MLN)
is defined by a quadruplet $M = (V_M, E_M, V^N, \bm{L})$, where the set of $N$
nodes is given by $V^N = \lbrace 1, 2, ..., N \rbrace$. The multidimensional
layer structure is given by a sequence of sets of elementary layers, $\bm{L} =
\lbrace L_a \rbrace^{d}_{a=1}$, where each of the $d$ sets of elementary layers
$L_a$ corresponds to an `aspect' $a$ of the MLN (e.g., $L_1 = \lbrace
\text{facebook}, \text{twitter}, ...  \rbrace$ could be a set of categories of
connections, and $L_2 = \lbrace \text{2010}, \text{2011}, ... \rbrace$ could be
a set of time stamps, at which edges are present). A layer in the structure
given by $\bm{L}$ is then a combination of elementary layers from all aspects,
or in other words: an element of the set of all layers given by the Cartesian
product $L_1 \times \cdot \cdot \cdot \times L_d$. Each node can belong to any
subset of the layers, and the set of all existing node-layer tuples (in short:
node-layers) $(u, \boldsymbol{\alpha})$, where $u \in V^N$ and
$\boldsymbol{\alpha} \in L_1 \times \cdot \cdot \cdot \times L_d$, is denoted
$V_M \subseteq V^N \times L_1 \cdot \cdot \cdot \times L_d$. Edges are allowed
between all such existing node-layers, hence the set of edges is given by $E_M
\subseteq V_M \times V_M$.

The pair $G_M = (V_M, E_M)$, referred to as the `supra-graph' of $M$, is a graph
on its own, where nodes are, as the authors say, ``labelled in a certain way''.
The adjacency matrix of $G_M$ is referred to as the `supra-adjacency matrix'
representation of $M$, and constitutes one possible representation of a MLN.
Defining weights for edges of $M$ on the underlying graph $G_M$ (by some
function $w: E_M \rightarrow \mathbb{R}$) yields a weighted MLN.

Another representation of a MLN can be achieved by adjacency
tensors~\cite{DeDomenico2013}. Given a MLN $M = (V_M, E_M, V^N, \bm{L})$ with
$d$ aspects, one can represent it by an order-$2(d+1)$ adjacency tensor
$A_{uv\boldsymbol{\alpha}\boldsymbol{\beta}} =
A_{uv\alpha_1\beta_1...\alpha_d\beta_d}$, where an element
$A_{uv\boldsymbol{\alpha}\boldsymbol{\beta}}$ has a value of $1$, if and only if
$((u, \boldsymbol{\alpha}), (v, \boldsymbol{\beta})) \in E_M$, and a value of
$0$ otherwise. As the authors of~\cite{kivela2014multilayer} explain, the
representation of a MLN by an adjacency tensor is technically only valid for
node-aligned MLNs, where all layers contain all nodes, $V_M = V^N \times L_1
\times \cdot \cdot \cdot \times L_d$. Yet, many tensor-based methods on MLNs
have been successfully applied by filling layers with `empty' node-layers
(node-layers that are not adjacent to any other node-layer), yielding an
artificial node-aligned structure of the MLN. However, one has to be very
cautious in the calculation and interpretation of tensor-based measures, and
account for the presence of empty node-layers in an appropriate
way~\cite{kivela2014multilayer}. In the tensor-representation of MLNs, weights
can be introduced by defining a weighted adjacency tensor
$W_{uv\boldsymbol{\alpha}\boldsymbol{\beta}}$, where the value of each element
determines the weight of an edge (for non-existing edges, the value is $0$ by
convention).

\section{Discussion of Multilayer Networks} \label{sec:Discussion of Multilayer
Networks}

In this section, we first demonstrate the alternative representation of a
multilayer network (MLN) by our framework, which is given by placing the
additional information attributed to the layered structure of a MLN $M$ in the
edges of $G = (V, E)$. Then, we show the advantages of the representation stated
in the main text. For that matter, we create the subset of the partition lattice
${}^GL$ of $G \mathrel{\widehat{=}} M$ that is induced by the types of features
of its constituent nodes, and show that it incorporates not only the alternative
representation shown here, but several others, including a tensor-like
representation. Lastly, we discuss the constraints imposed on our framework in
order to represent a MLN, and explain how our framework solves the issues
encountered in the representation of MLNs.

The alternative representation of $M = (V_M, E_M, V^N, \bm{L})$ by $G = (V, E)$
is given by identifying each node $V_i = \lbrace i \rbrace \in V = \lbrace V_1,
V_2, ..., V_N \rbrace$ with a node $V^N_i \in V^N$, $V_i \mathrel{\widehat{=}}
V^N_i$. Denoting the weight of an edge of a MLN by $w \left( ( (V^N_i,
\boldsymbol{\alpha}),(V^N_j, \boldsymbol{\beta}) ) \right) \in \mathbb{R}$, an
edge $E_{ij} \in E' = \lbrace E_{11}, E_{12}, ..., E_{NN} \rbrace$ is given by
\begin{equation}
E_{ij} = \lbrace w \left( (
(V^N_i,\boldsymbol{\alpha}),(V^N_j,\boldsymbol{\beta}) ) \right) \;\vline\;
((V^N_i,\boldsymbol{\alpha}), (V^N_j,\boldsymbol{\beta})) \in E_M \rbrace =:
\lbrace R^k_{ij} \;\vline\; k \in \lbrace 1, 2, ..., r_{ij} \rbrace \rbrace,
\end{equation}
where $|E_{ij}| = r_{ij}$ is the number of types of relations from node $V_i$ to
node $V_j$. Hence, the edge set $E$ corresponding to $E_M$ is given by $E =
\lbrace E_{ij} \;\vline\; i,j \in \lbrace 1, 2, ..., N \rbrace \land E_{ij} \neq
\varnothing \rbrace$. By this representation, we can clearly see that a tuple
$(\boldsymbol{\alpha},\boldsymbol{\beta})$ defines the type of relation of an
edge in $E_M$,
\begin{equation}
t \left( ((V^N_i,
\boldsymbol{\alpha}),(V^N_j,\boldsymbol{\beta})) \right) = t \left( ((V^N_k,
\boldsymbol{\gamma}),(V^N_l,\boldsymbol{\delta})) \right) \longleftrightarrow
(\boldsymbol{\alpha}, \boldsymbol{\beta}) = (\boldsymbol{\gamma},
\boldsymbol{\delta}),
\end{equation}
for all $V^N_i,V^N_j,V^N_k,V^N_l \in V^N$, where $t$ is
a function mapping an edge to its corresponding type, $t: E_M \rightarrow T =
\lbrace 1, 2, ..., m_{\text{types}} \rbrace$, with $m_{\text{types}} =
(\prod^{d}_{a=1}|L_a|)^2$. Therefore, the number of types of relations between
any pair of nodes in a MLN is bounded by $r_{ij} \leq m_{\text{types}}$.

Next, we partition the graph $G = (V, E) \mathrel{\widehat{=}} (V_M, E_M)$
described by Eqs.~(\ref{eq:mln1}) and (\ref{eq:mln2}). For notational
uniformity, we rewrite the features of the nodes in $V$ as outputs of partition
functions $p = \lbrace p^N, p^1, p^2, ..., p^d \rbrace$, where
\begin{align}
&p^N: V \rightarrow V^N,  V_i \mapsto p^N(V_i) = V^N_i, \\
&p^a: V \rightarrow L_a,  V_i \mapsto p^a(V_i) = L_{a,i},  a = 1, 2, ..., d.
\end{align}
Based on the $(1+d)$ partitions induced by $p$, we can redistribute the
information contained in the graph $G$ on a subset of the lattice ${}^GL^f
\subseteq {}^GL$. This redistribution allows for several representations of the
graph $G$, some of which we will demonstrate in the following. Let us denote the
partition index set of $p$ by $I^K = \lbrace N, 1, 2, ..., d \rbrace$. Then we
can select a total of $I(K) = 2^{(1+d)}$ distinct collections $g \subseteq I^K$,
resulting in $|{}^GL^f| \leq I(K)$ supergraphs $G^{\underline{p}} =
(V^{\underline{p}}, E^{\underline{p}}) \in {}^GL^f$, where ${}^GL^f = \lbrace
G^{\underline{p}} \;\vline\; g \in \mathcal{P}(I^K) \rbrace$ and $\underline{p}
= (p^k)_{k \in g}$.

Choosing $g = \lbrace N \rbrace$ leads to the supergraph $G^{p^N} = (V^{p^N},
E^{p^N})$, where each supernode $V^{p^N}_i \in V^{p^N}$ corresponds to a node of
the MLN, $V^{p^N}_i \mathrel{\widehat{=}} V^N_i$. Superedges $E^{p^N}_{ij} \in
E^{p^N}$ with $i = j$ correspond to the coupling edges of a MLN. The one to one
correspondence of the supergraph $G^{p^N}$ to the above, edge-based choice of
$G$ justifies the statement that the representation $G$ of $M$ given in the main
text is the preferred one, since it fully entails the above choice.

Choosing the group $g = \lbrace 1, 2, ..., d \rbrace$ leads to the supergraph
$G^{p^1 \cdot\cdot\cdot p^d} = (V^{p^1 \cdot\cdot\cdot p^d},E^{p^1
\cdot\cdot\cdot p^d})$, where every supernode $V^{p^1 \cdot\cdot\cdot p^d}_{i^1
\cdot\cdot\cdot i^d} \in V^{p^1 \cdot\cdot\cdot p^d}$ corresponds to a distinct
layer of $M$, encompassing all its respective nodes. Superedges $E^{p^1
\cdot\cdot\cdot p^d}_{i^1 \cdot\cdot\cdot i^d, j^1 \cdot\cdot\cdot j^d} \in
E^{p^1 \cdot\cdot\cdot p^d}$ with either $(i^a)_{a=1}^d = (j^a)_{a=1}^d$ or
$(i^a)_{a=1}^d \neq (j^a)_{a=1}^d$ correspond to collections of intra- and
inter-layer edges of a MLN, respectively.

The last supergraph we want to exemplify is given by choosing $g = \lbrace N, 1,
2, ..., d \rbrace = I^K$, resulting in the supergraph $G^{p^N \cdot p^1
\cdot\cdot\cdot p^d} = (V^{p^N \cdot p^1 \cdot\cdot\cdot p^d},E^{p^N \cdot p^1
\cdot\cdot\cdot p^d})$. This supergraph corresponds one to one to the graph $G =
(V, E)$, and therefore to the `supra-graph' representation of $M$, given by the
tuple $(V_M, E_M)$. The only difference is the indexing. The graph $G$ has an
adjacency matrix-like representation, given by $E_{ij} \in E'$. We say `like',
since $E'$ is not a matrix, formally. An element of $E'$ is either a real
number, corresponding to the weight of the corresponding edge in $E_M$, or an
empty set, meaning the edge does not exist. $G^{p^N \cdot p^1 \cdot\cdot\cdot
p^d}$, on the other hand, has a tensor-like representation, given by $E^{p^N
\cdot p^1 \cdot\cdot\cdot p^d}_{i^N \cdot i^1 \cdot\cdot\cdot i^d, j^N \cdot j^1
\cdot\cdot\cdot j^d} \in E^{p^N \cdot p^1 \cdot\cdot\cdot p^d}$. Again,
formally, $E^{p^N \cdot p^1 \cdot\cdot\cdot p^d}$ is not a tensor. An element of
$E^{p^N \cdot p^1 \cdot\cdot\cdot p^d}$ is either a real number, corresponding
to the weight of the corresponding edge in $E_M$, or an empty set, if the edge
does not exist. As mentioned in Sec.~\ref{sec:Intersection Partitions}, we can
distinguish between a superedge that does not exist because at least one of the
supernodes does not exist, $n^{p^N \cdot p^1 \cdot\cdot\cdot p^d, i^N \cdot i^1
\cdot\cdot\cdot i^d}$ or $n^{p^N \cdot p^1 \cdot\cdot\cdot p^d,j^N \cdot j^1
\cdot\cdot\cdot j^d} = 0$, or because there is no superedge between existing
supernodes, $n^{p^N \cdot p^1 \cdot\cdot\cdot p^d, i^N \cdot i^1 \cdot\cdot\cdot
i^d}$ and $n^{p^N \cdot p^1 \cdot\cdot\cdot p^d,j^N \cdot j^1 \cdot\cdot\cdot
j^d} = 1$.

From the perspective of our framework, all representations $G^{\underline{p}}
\in {}^GL$ are equivalent, in the sense that the information contained in $G$ is
conserved under partitioning. There is no need to ``flatten'' the MLN
represented by $G^{p^N \cdot p^1 \cdot\cdot\cdot p^d}$ to obtain its
supra-adjacency matrix representation $G$, and there is no loss of information
about the aspects, as -- according to~\cite{kivela2014multilayer} -- it is the
case for MLNs represented by $M = (V_M, E_M, V^N, \bm{L})$.

Let us now summarize the constraints we imposed on our framework, in order to
represent a MLN. First, we had to restrict ourselves to the representation of
one element of a deep graph. Allocating information on and between groups of
nodes, as described in Sec.~\ref{sec:Redistribution and Allocation of
Information on the Lattices}, is not intended within the framework of MLNs.
Then, we have to decide whether to put to information attributed to the layered
structure of $M$ into the nodes of $G$, or the edges of $G$. There is no genuine
separation of features and relations in a MLN. Furthermore, the weights
of the edges of a MLN need to be restricted to real numbers (or possibly
complex numbers). This poses several limitations.
First, it is problematic to distinguish between edges with a weight of $0$ (e.g.
an edge representing a time difference of $0$) and non-existing edges, since
edges with weight $0$ do not exist by convention in MLNs. Yet, more importantly,
we can not assign distributions of values to nodes or edges, let alone more
complex mathematical objects. Another complication arises, when dealing with
nodes that have more or less than $d$ aspects, or more generally speaking: when
dealing with heterogeneous kinds of nodes. Although it is possible to represent
nodes with different types of features by filling layers with `empty'
node-layers, the procedure is rather counter-intuitive and leads to a cluttered
representation. In contrast, our framework provides the means to represent
heterogeneous objects and their relations in a sparse and intuitive manner.

\rowcolors{2}{gray!25}{white}
\begin{table*}
\caption{\label{tab:glossary}\textsf{Glossary.} The symbol
``\#'' reads: ``number of'', and ``IP'' reads: ``intersection partition''.}
\begin{ruledtabular}
\begin{tabular}{lccl}
\rowcolor{gray!50}
Explanation & Symbol & Given by & Properties \\
\hline
\hline
\# nodes & $n$ & $|V|$ & $\geq 1$ \\
\hline
\# supernodes & $n^p$ & $|V^p|$ & $1 \leq n^p \leq n$ \\
\hline
\# supernodes (IP) & $n^{\underline{p}}$ & $|V^{\underline{p}}|$ &
$1 \leq n^{\underline{p}} \leq n$ \\
\hline
\# nodes in supernode $i$ & $n^{p,i}$ & $|V^p_i|$ & $1 \leq n^{p,i} \leq n$  \\
\hline
\# nodes in supernode $\underline{i}$ (IP) & $n^{\underline{p}, \underline{i}}$
& $|V^{\underline{p}}_{\underline{i}}|$ & $0 \leq n^{\underline{p},
\underline{i}} \leq n$ \\
\hline
\# types of features in supernode $\underline{i}$ (IP) & $n^{\underline{p},
\underline{i}}_{\text{types}}$ & $|T^{\underline{p}}_{\underline{i}}|$ & $0
\leq n^{\underline{p}, \underline{i}}_{\text{types}} \leq
|F^{\underline{p}}_{\underline{i}}|$ \\
\hline
\# features of type $t$ in supernode $\underline{i}$ (IP) & $n^{\underline{p},
\underline{i}}_t$ & $|F^{\underline{p}, T}_{\underline{i}, t}|$ & $\leq
n^{\underline{p}, \underline{i}}$ \\
\hline
\# partition-specific types of features in supernode $\underline{i}$ (IP) &
${}^{\underline{p}}n^{\underline{p}, \underline{i}}_{\text{types}}$ & allocation
& $\geq 0$ \\
\hline
\# features in node $i$ & $f_i$ & $|V_i|$ & $\geq 0$ \\
\hline
\# distinct types of features in $G$ & $n_{\text{types}}$ & $|T_v|$ &
$\geq 0$ \\
\hline
\# features of type $t$ in $G$ & $n_t$ & $|F^{p^c, T}_{i, t}|$ & $\leq n$ \\
\hline
\hline
\# edges & $m$ & $|E|$ & $\geq 0$ \\
\hline
\# superedges & $m^p$ & $|E^p|$ & $0 \leq m^p \leq m$ \\
\hline
\# superedges (IP) & $m^{\underline{p}}$ & $|E^{\underline{p}}|$ & $0 \leq
m^{\underline{p}} \leq m$ \\
\hline
\# edges in superedge $(i,j)$ & $m^{p,ij}$ & $|E^p_{ij}|$ & $0 \leq m^{p,ij}
\leq m$ \\
\hline
\# edges in superedge $(\underline{i},\underline{j},\underline{r})$ (IP) &
$m^{\underline{p},\underline{i}\underline{j}, \underline{r}}$ &
$|E^{\underline{p}}_{\underline{i}\underline{j}, \underline{r}}|$ & $0 \leq
m^{\underline{p},\underline{i}\underline{j}, \underline{r}} \leq m$ \\
\hline
\# types of relations in superedge $(\underline{i}, \underline{j},
\underline{r})$ (IP) & $m^{\underline{p},\underline{i}\underline{j},
\underline{r}}_{\text{types}}$ & $|T^{\underline{p}}_{\underline{i}
\underline{j}, \underline{r}}|$ & $0 \leq m^{\underline{p},\underline{i}
\underline{j}, \underline{r}}_{\text{types}} \leq |R^{\underline{p}}_{
\underline{i}\underline{j}, \underline{r}}|$ \\
\hline
\# relations of type $t$ in supernode $(\underline{i},\underline{j},
\underline{r})$ (IP) & $m^{\underline{p},\underline{i}\underline{j},
\underline{r}}_t$ & $|R^{\underline{p}, T}_{\underline{i}\underline{j},
\underline{r}, t}|$ & $\leq m^{\underline{p},\underline{i}\underline{j},
\underline{r}}$ \\
\hline
\# partition-specific types of relations in superedge $(\underline{i},
\underline{j}, \underline{r})$ (IP) & ${}^{\underline{p}}m^{\underline{p},
\underline{i}\underline{j}, \underline{r}}_{\text{types}}$ & allocation &
$\geq 0$ \\
\hline
\# relations in edge $(i,j)$ & $r_{ij}$ & $|E_{ij}|$ & $\geq 0$ \\
\hline
\# distinct types of relations in $G$ & $m_{\text{types}}$ & $|R^{p^c}_{ij}|$ &
$\geq 0$ \\
\hline
\# relations of type $t$ in $G$ & $m_t$ & $|R^{p^c, T}_{ij, t}|$ & $\leq m$
\end{tabular}
\end{ruledtabular}
\end{table*}

\begin{acknowledgments}

This paper was developed within the scope of the IRTG 1740/TRP 2011/50151-0,
funded by the DFG/FAPESP. NB acknowledges financial support by the Alexander von
Humboldt Foundation and the German Federal Ministry for Education and Research.

\end{acknowledgments}

\bibliography{deepgraph}

\end{document}